\documentclass[prb,11pt]{revtex4-1}

\usepackage{amsmath}    
\usepackage{graphicx}   
\usepackage{verbatim}   
\usepackage{color}      
\usepackage{subfigure}  
\usepackage{hyperref}   
\newcommand{\mb}{\mathbf}
\newcommand{\Real}{\mathrm{Re\,}}
\newcommand{\Imag}{\mathrm{Im\,}}
\graphicspath{{./}, {./figs/}}
\begin{document}

\title{From active stresses and forces to self propulsion of droplets}
\author{ R. Kree,$^\ast$ P.S. Burada,$^\dagger$ and A. Zippelius$^\ast$}
\affiliation{$^\ast$Georg-August-Universit\"at G\"ottingen, 
Institut f\"ur Theoretische Physik, Friedrich-Hund-Platz 1, 37077 G\"ottingen, Germany\\ 
$^\dagger$ Department of Physics, Indian Institute of Technology, Kharagpur -- 721302, India}



\date{\today}

\begin{abstract} 
We study the self-propulsion of spherical droplets as simplified hydrodynamic models of swimming microorganisms or artificial microswimmers.  In contrast to approaches, which start from active velocity fields produced by the system, we consider active surface force or body force densities or active stresses as the origin of autonomous swimming. For negligible Reynolds number and given activity we first calculate the external and the internal flow fields as well as the center of mass velocity and an angular velocity of the droplet at fixed time.  To construct trajectories from single time snapshots, the evolution of active forces or stresses must be determined in the laboratory frame. Here, we consider the case of active matter, which is carried by a continuously distributed, rigid but sparse (cyto)-sceleton that is immersed in the droplet's interior. We calculate examples of trajectories of a droplet and its sceleton from force densities or stresses, which may be explicitely time dependent in a frame fixed within the sceleton.  
\end{abstract}
\maketitle

\section{\label{introduction}Introduction}
Ever since the early work of \cite{Lighthill_1952} and
others \cite{Blake_1971,Purcell_1977}, the self-propulsion of
microorganisms has attracted a lot of
interest \cite[for reviews see]{Lighthill_1976,Lauga_2009}. Over the last ten years or
so, the activity increased enormously, partially motivated by the
successful design of artificial swimmers \cite{Ebbens_2010}. Both,
biological as well as artificial swimmers have been studied
extensively \cite{Lauga_2009}, experimentally and with help of
analytical modelling and simulation.

In the world of living microorganisms, there are many specific
examples of self-propelled motion. One possibility are controlled
shape changes of the organism as the cause of self-propulsion. These
changes may either occur on the scale of the whole system, like in
amoeboid or flagella-driven motion, or they may be confined to a thin
boundary layer at the interface, like in ciliar motion of
microorganisms. In the so called squirmer models, the small and
periodic tangential shape changes of a quasi-continuous envelope of
cilia drive a flow field close to the interface, which, after
time-averaging, is modelled as a boundary condition for the velocity
field at the interface \cite{Blake_1971}. 
Some microorganisms, as for
example the cyanobacterium \textit{Synechococcus} or \textit{Myxobacteria} do
not even possess external organelles associated with motility, but
seem to be driven by helical mechanical waves of the cell membrane \cite{EhlersOster2012, Nanetal2011}.

On the other side, many artificial systems have been fabricated,
allowing for high control of the mechanisms of self propulsion and the
resulting dynamics of single swimmers as well as collective
motion \cite{Bechinger2016}. Common mechanisms are based on chemical
reactions on the surface of the
swimmer \cite{Golestanian2005,Takabatake2011,Herminghaus2012,Herminghaus2013}
and more generally phoretic effects \cite{Golestanian2007,Herminghaus2013}.

In addition, biomimetic devices have been assembled
and analysed, such as solutions of cytoskeletal filaments and
motors \cite{Sanchez2012,Koehler2011}.

All the processes, which produce active flow and self-propulsion of
microorganisms are controlled from the interior of the cell, in
particular from the cytosceleton. Most frequently studied models, such
as squirmer models, which use active flow boundary conditions at the
interface, do not specify the forces inside the organism which provide
a given squirming motion. On a molecular level, detailed knowledge
about the molecular machinery is required to answer the question which
forces inside the droplet provide a given squirming motion (or any
other mechanism of self-propulsion).  But on a mesoscopic,
hydrodynamic level, the question may be answered in a more universal
way.  The internal mechanical properties of cells have been modelled
as a material continuum \cite{Dembo1986}, with a distribution of active
force densities or stress tensors \cite{Joanny2007}, which are
produced by molecular motors and (dis)-assembly of semi-flexible
polymers \cite{Loisel1999}. A complete mesoscopic mechanical model of
self-propulsion should consider the coupled evolution of active and
passive matter and relate it to the motion of the whole cell and the
generated flow fields in the interior of the cell and in the ambient
fluid.

Here, we want to provide such a unified analytic framework, which
links trajectories and flow fields of a swimmer to active forces. The
swimmer is modeled as a droplet, consisting of two immiscible
fluids. Activity is generated by one or several of the following
causes: (i) active surface force densities (tractions) $\mb{t}^{act}$
on the interface $\Gamma(t)$ , (ii) active body force densities
$\mb{f}^{act}$ or (iii) active stresses $\pmb{\sigma}^{act}$ in the
interior, $K(t)$, of the droplet .
Each of the active force distributions is the outcome
of a molecular machinery, which we do not model in detail.  Interface
tractions may arise from the Marangoni effect, i.e. an inhomogeneous
distribution of surfactants in $\Gamma$ leading to inhomogeneous
surface tension. Another machinery to generate surface tractions is
made of molecular motors located at $\Gamma$, which drive short
protrusions in the ambient fluid. These tractions only appear in the
boundary conditions at $\Gamma$, which will be specified below.
Internal active stresses emerge from a continuum description of short
range molecular interactions (for example in the context of an active
polar medium) inside the droplet and thus vanish outside of $K(t)$ by
definition.  They may generate body force densities
$\mb{f}^{act}=-\nabla\cdot\pmb{\sigma}^{act}$, interface tractions
$-\mb{e}_r\cdot\pmb{\sigma}^{act}|_\Gamma$ or both.
An active body
force density may, for example, be the result of a distribution of
molecular motors driving objects (molecular paddles) in the droplet's
interior. 
To keep the coupled dynamics of internal material flows simple, we
introduce \textit{rigid active matter}, i.e. a rigid polymeric
structure ---modelling a (cyto)-sceleton --- which we describe by a
continuous density carrying active forces.

Surface tractions, active body forces or active stresses in the
interior generate flow fields both outside and inside the droplet,
which can indeed be observed \cite{Drescher2010, Mustacich1976}. We
compute these flow fields as well as the resulting trajectory of the
droplet.  Thus it becomes possible to infer, which actively generated
internal flows or which active forces in a mechanical cortex near the
interface may cause a given flow field and droplet motion in the
ambient fluid, and \textit{vice versa}.  The analysis of general
surface tractions also includes the Marangoni mechanism and the
squirmer model as special cases.  We discuss autonomous swimming which
has to respect the constraints that total force and torque on the
droplet vanish. Beyond autonomous swimming, we also analyse simple
cases of the droplet`s motion in external fields, such as
sedimentation and buoyancy or optical traps.

The paper is organized as follows: In sec.\ref{model} we introduce the
model, specify the boundary conditions and set up the general solution strategy. In
sec.\ref{activity}, we discuss flow fields of surface tractions, connect the
force- and torque free case to the squirmer, and show that forces
which can be derived from a potential are equivalent to given surface
tractions. In sec.\ref{bodyforces}, we discuss a general class of
active force densities in the interior of the droplet and in
sec.\ref{polarmedium} we consider the case of active stresses,
which are generated by a polar medium. All calculations and results of sec. \ref{activity}-\ref{polarmedium} refer to a single instant in time and therefore need no evolution equation of the active material.  In
sec.\ref{sec:trajectories}, we discuss the coupled dynamics of rigid active matter and internal flow and discuss some illustrating cases of emerging trajectories, before ending with conclusions and an outlook in
sec.\ref{conclusions}. Details of the calculations are delegated
to Appendices.

\section{Model}\label{model}
Embedded within a Newtonian fluid consider a droplet of another
viscous fluid, which at time $t$ occupies a finite region $K(t)$ in
3-dimensional space.  The two fluids differ in viscosity and are
assumed to be incompressible and completely immiscible. Inside the
droplet or located at its geometric boundary $\Gamma(t)=\partial K(t)$
we assume active force densities, and we want to study, how they lead
to self-propulsion of the droplet in a regime of negligible Reynolds
number \cite{Happel, KimKarrila2005}.  In a first step, we consider the forces as given. In a second
step, which will be discussed in detail in
sec. \ref{sec:trajectories}, we study the coupled dynamics of flow and
active forces carried by a rigid (cyto)-sceleton within the droplet.
We restrict the class of systems under study to spherical droplets (of
fixed radius $a$), which do not change their shape during time
evolution. The field of normal unit vectors on $\Gamma$ thus consists
of radial unit vectors $\mb{e}_r$ pointing into the ambient fluid.

At every fixed instant of time, the flow velocity fields $\mathbf{v}^\pm$ inside ($-$) and outside ($+$) the droplet have to obey the Stokes equations in the laboratory frame (LF),
\begin{equation}
	\eta^{\pm}\nabla^2 \mathbf{v}^{\pm}-\nabla p^\pm=\nabla\cdot\pmb{\sigma}^\pm=-\mb{f}^{act}-\mb{f}^{ext} , \quad \nabla\cdot \mathbf{v}^{\pm}=0,
\label{eq:stokes}
\end{equation} 
with viscosity $\eta^\pm$ inside and outside the droplet. 
The flow fields $\mb{v}^\pm$ create viscous stresses
$\pmb{\sigma}^{\pm}$ with cartesian components
$\sigma^{\pm}_{ij}=-p^{\pm}\delta_{ij}+\eta^{\pm}(\partial_jv^{\pm}_i+\partial_iv^{\pm}_j)$,
where $\partial_iv_j$ denotes the partial derivative
$\partial v_j/\partial x_i$.  The pressure fields $p^{\pm}$ are fixed
by the incompressibility condition up to a constant.  The active force
density, $\mb{f}^{act}$, vanishes outside of the droplet.  We can
either prescribe the force density or an active stress, which uniquely
determines $\mb{f}^{act}=-\nabla\cdot\pmb{\sigma}^{act}$.  Note,
however, that the reverse is not true. Generally it is not possible to
calculate a unique stress from a body force density by solving
$-\nabla\cdot\pmb{\sigma}^{act}=\mb{f}^{act}$ without any further
information about the underlying molecular interactions. For the
purposes of our general framework, we therefore treat the problems of
given $\mb{f}^{act}$ and given $\pmb{\sigma}^{act}$ separately.

To determine $\mb{v}^\pm$ uniquely, the Stokes equations have to be supplemented by boundary conditions at
the interface and at infinity. 
There are two boundary conditions at $\Gamma(t)$.  First, we use the
assumption that there is no velocity slip, i.e. the tangential
velocity component is continuous across the interface. In combination
with the condition of immiscibility, this leads to the continuity of
$\mb{v}$ across $\Gamma(t)$,
\begin{equation}
\label{eq:continuity_v}
	\mb{v}^+(\mathbf{r}+\epsilon\mb{e}_r)-\mb{v}^-(\mb{r}-\epsilon\mb{e}_r)=\mb{0}\quad \text{for}\; \epsilon\to 0,\;  \mb{r}\in\Gamma(t)
\end{equation}     

The second boundary condition at $\Gamma(t)$ simply states that the
interface is free, i.e. all forces acting on each area element must
add up to zero. 
The viscous
tractions $\mb{t}^{\pm}=\pm\mb{e}_r\cdot\pmb{\sigma}^{\pm}$, which act on 
$d^2\mb{S}=d^2S\,\mb{e}_r$ at the interface  are balanced by a variety of forces,
from which we separate a ubiquitous Laplace pressure term
$\mb{t}_L=-\gamma_0(\nabla\cdot\mb{e}_r)\mb{e}_r=-(2\gamma_0/a)\mb{e}_r$
resulting from a homogeneous surface tension $\gamma_0$, so that
$\mb{t}^+ +\mb{t}^- +\mb{t}_L+\mb{t}_{a}=0$ on
$\Gamma(t)$.  If the active stress, $\pmb{\sigma}^{act}$ is nonvanishing at the interface, it will
also contribute to $\mb{t}_a$. 
We separate this term explicitly and write
$\mb{t}_a=- \mb{e}_r\cdot\pmb{\sigma}^{act}|_\Gamma+{\mb{t}}^{act}$. The
boundary condition then takes on the form
\begin{equation}
\mb{e}_r\cdot(\pmb{\sigma}^+ - \pmb{\sigma}^-)
        =\frac{2\gamma_0}{a}\mb{e}_r+\mb{e}_r\cdot\pmb{\sigma}^{act}-\mb{t}^{act}\quad \text{on}\; \Gamma(t).
	\label{eq:continuity_trac}
\end{equation}

We do not consider any ambient background flow in LF; hence the
velocity should vanish for $|\mb{r}|\to\infty$, and the total force
acting on the surface of an infinitely large sphere should stay
finite. This requires $\mb{v}^+(\mb{r})={\cal O}(1/|\mb{r}|)$ for
$|\mb{r}|\to\infty$. The pressure is only determined up to a constant
reference, which we fix by $p^+(\mb{r})\to 0$ for
$|\mb{r}|\to\infty$. For the autonomous swimmer one furthermore
requires that it is force free, so that momentum is conserved. 
Therefore the momentum flux
$\mb{J}^+(r)=\int_r d^2S\; \pmb{\sigma}^+\cdot\mb{e}_r$ through the
surface of a sphere of radius $r$ vanishes for ${r\to\infty}$. The viscous stresses thus have to vanish faster
than $r^{-2}$, so that the flow velocity has to vanish faster than
$r^{-1}$. This rules out the so called Stokeslet solutions with
$\mb{v}^+\propto r^{-1}$ for ${r\to\infty}$.
For
active tractions and body forces, the requirement
\begin{equation}\label{eq:totaltractionforce}
\mb{F}=\int_{\partial V} d^2S \; \mb{t}^{act} + \int_{V} dV \;\mb{f}^{act}=0 
\end{equation}
constitutes an additional constraint. 
This constraint is fulfilled automatically, if
$\mb{f}^{act}$ and $\mb{t}^{act}$ result from active stresses , because 
 $ \int_V dV\, \nabla\cdot\pmb{\sigma}^{act}=0 $ due to
Gauss' theorem for any $V$ enclosing the droplet.

 Similar arguments hold
for the torques. For autonomous swimmers, the total torque has to
vanish, and the requirement
\begin{equation}\label{surface-torque-free}
\mb{N}=\int_{\partial V} d^2S \;\mb{r}\times\mb{t}^{act} + \int_{V} dV \;\mb{r}\times\mb{f}^{act}=0
\end{equation}
is an additional constraint on $\mb{t}^{act}$ and $\mb{f}^{act}$. The torque generated by active stresses vanishes automatically, if $\pmb{\sigma}^{act}$ is symmetric.

If additional external fields, such as buoyancy due to density
mismatch, walls in a microfluidic device or optical traps are present,
the total momentum is no longer conserved and the momentum sources or
sinks resulting from external force densities $\mb{f}_{ext}(\mb{r},t)$
have to be added to the above relations. These force densities are
usually specified as functions in the laboratory frame.

The translational motion of the  droplet is characterized by the velocity $\mb{v}_{CM}$ of its center of mass.
For an incompressible flow $v_i=\partial(v_ix_j)/\partial x_i$ and consequently,   $\mb{v}_{CM}$ can alternatively be expressed as an integral over the interface,
\begin{equation}
\label{vCM}
	\mb{v}_{CM}(t)=\frac{3}{4\pi a^3}\int_K \mb{v}^-\, dV=\frac{3}{4\pi a^2}\int_{\Gamma(t)} (\mb{v}^\pm\cdot\mb{e}_r)\,\mb{e}_r d^2S.  
\end{equation}
This equation simply expresses the fact that the total momentum of the droplet equals the velocity of its center of mass, multiplied by its total mass. An angular velocity may be defined analogously as the total angular momentum divided by the moment of inertia $I=8\pi a^5/15$ of the homogeneous spherical droplet:
\begin{equation}
	\label{angular-momentum}
	\pmb{\omega}=\frac{1}{I}\int_K \mb{r}\times\mb{v}^-\, dV.
	\end{equation}	
The significance of $\pmb{\omega}$ for the trajectories of the droplet will be discussed in Sec. \ref{sec:trajectories}. 

Active tractions, body force densities and stresses will in general drive an
active flow velocity field $\mb{v}_{act}$ inside the droplet and on
$\Gamma(t)$. This field  provides the link
between models based on force densities and models based on active
velocities, in particular the envelope approximation of squirmer
models. We will discuss this link in sec.~\ref{squirmer-droplet}.  

Our general strategy to study self-propulsion consists of two steps. First,
the boundary value problem posed by eqs. (\ref{eq:stokes}, \ref{eq:continuity_v}, \ref{eq:continuity_trac}) is solved
at a fixed instant of time by expanding all quantities in terms of scalar or vector spherical harmonics, each characterized by a set of integer numbers $0\leq \ell < \infty$ and $-\ell\leq m \leq \ell$. The definitions of the vector spherical harmonics and details of
the solution procedure can be found in Appendix A. Stokes equations become a system of uncoupled ordinary differential equations for the  mode amplitudes. Rigid body motions of the droplet are exclusively contained in the $\ell=1$ modes, which, for self-propelling droplets, contribute a term $\mathcal{O}(r^{-3})$ to $\mb{v}^+$.  
The leading contributions  $\mb{v}^+=\mathcal{O}(r^{-2})$ for large $r$ emerge from  $\ell=2$ modes, and some of the $\ell=3$ modes also decay as $\mathcal{O}(r^{-3})$. Thus reduced descriptions of the force densities are sufficient, if one is only interested in the trajectories of the droplets and the far field of the accompanying flow. In the present work, we will focus on the droplet's motion. Therefore, 
in a second step, we set up the temporal evolution of active forces and construct the common translational motion of the droplet and the rigid sceleton and the rotational motions of these two constituents.

\section{Activity generated by traction forces}
\label{activity}
In this section, we consider a droplet with given active tractions,
$ {\bf t}^{act}$ on its surface. It is our goal to compute the
external, ${\bf v}^+$, and internal, ${\bf v}^-$, flow fields as well
as the center of mass velocity, ${\bf v}_{CM}$, and the rotational
velocity, $\pmb{\omega}$, of the droplet.  We first consider general
tractions, exerting a nonzero force and torque on the droplet.
Subsequently, we specialise to the force- and torque-free case,
i.e. discuss an autonomous swimmer which is propelled by surface
tractions. We compare our results to the well studied squirmer with an
active flow prescribed on the surface. The linear velocity as well as
the external flow can be mapped onto each other, whereas the
rotational velocity of the droplet driven by tractions vanishes. The
droplet furthermore develops a characteristic flow pattern in the
interior and energy is dissipated by both, the interior and exterior
flow with the ratio of the two determined by the corresponding ratio
of viscosities.  We also show that external forces which can be
derived from a potential, such as gravity, can be mapped onto
tractions. The motion of the droplet with active tractions and
external force is worked out.

\subsection{ General tractions}
\label{tractions}
To determine the droplet's trajectories we start from the $\ell=1$ modes of active tractions in the MMF at a fixed time $t$, which we write as 
\begin{equation}
\label{eq:general_traction}
\mb{t}^{act}_{(\ell=1)}(\theta, \varphi)=\sum_{m=-1}^1\big(\alpha_{m}Y_{1m}(\theta, \varphi)\mb{e}_r
+\beta_{m}{\bf \nabla}_sY_{1m}(\theta, \varphi)
+ \gamma_{m} {\bf e}_r\times {\bf \nabla}_sY_{1m}(\theta, \varphi)
\big).
\end{equation}
Here $\nabla_s=r\nabla$ denotes the angular part of the gradient operator. 
We solve Stokes equations with the boundary conditions of
Eqs.(\ref{eq:continuity_v},\ref{eq:continuity_trac}) in the interior and exterior
region of the droplet and find for the velocity fields in the lab frame
\begin{equation}
{\bf v}^+({\bf r})=\sum_{m}\big((a_m^+\frac{a}{r}+b_m^+\frac{a^3}{r^3})
Y_{1m}(\theta, \varphi)\mb{e}_r
+(a_m^+\frac{a}{r}-b_m^+\frac{a^3}{r^3})\frac{1}{2}{\bf \nabla}_sY_{1m}(\theta, \varphi)+c_m^+\frac{a^2}{r^2}\mb{e}_r\times {\bf \nabla}_sY_{1m}(\theta, \varphi)
\big),\nonumber
\label{eq:vplus}
\end{equation}
\begin{equation}
{\bf v}^-({\bf r})=\sum_{m}\big((a_m^-\frac{r^2}{a^2}+b_m^-)
Y_{1m}(\theta, \varphi)\mb{e}_r
+(2a_m^-\frac{r^2}{a^2}+b_m^-){\bf \nabla}_sY_{1m}(\theta, \varphi)
+c_m^-\frac{r}{a}\mb{e}_r\times {\bf \nabla}_sY_{1m}(\theta, \varphi)
\big).
\label{eq:vminus}
\end{equation}
where the expansion coefficients are given in terms of the traction
forces as follows:
\begin{eqnarray}
a_m^+&=&\frac{a}{3\eta^+}(\alpha_m+2\beta_m)\\
b_m^+&=& - \frac{a}{3\eta^+}\frac{\lambda\alpha_m+2(\lambda+1)\beta_m}{2+3\lambda}\\
a_m^-&=& -\frac{a}{3\eta^+}\frac{\alpha_m-\beta_m}{2+3\lambda}\label{eq:amminus}\\
b_m^-&=&-\frac{a}{3\eta^+}\frac{(3+2\lambda)\alpha_m+(4\lambda+1)\beta_m}{2+3\lambda}\\
c_m^+&=&c_m^-=\frac{a\gamma_m}{3\eta^+}
\label{eq:bmminus}
\end{eqnarray}
Here the ratio of interior to exterior viscosity has been denoted by 
$\lambda=\eta^-/\eta^+$.

The above equations decouple into two sets of equations. The first one
corresponds to tractions with $\gamma_m=0$ that give rise to a linear
velocity of the droplet but no rotational velocity.
The center of mass velocity is determined by the flow field according to 
Eqs.\ref{vCM}.
Straightforward computation yields
\begin{equation}
\label{vCM_tractions1}
\mb{v}_{CM}=\sqrt{\frac{3}{4\pi}}h_0\mb{e}_z
-\sqrt{\frac{3}{2\pi}}\Real(h_1)\mb{e}_x
+\sqrt{\frac{3}{2\pi}}\Imag(h_1)\mb{e}_y\\
\end{equation}
with
\begin{equation}
\label{vCM_tractions2}
h_m=\frac{2a}{3\eta^+}
\frac{\alpha_m(1+\lambda)+\beta_m(1+2\lambda)}{2+3\lambda}
\end{equation}
Such tractions can for example be generated by Marangoni flow, which
is represented by an inhomogeneous surface tension
$\Gamma(\theta,\varphi)$. For this example the tractions are given by
\begin{equation}
a\mb{t}^{act}(\theta, \varphi)=2\Gamma(\theta,\varphi)\mb{e}_r
+{\bf \nabla}_s\Gamma(\theta,\varphi),
\end{equation}
and hence cannot give rise to a rotational velocity of the droplet.

The second class of tractions has only $\gamma_m\neq 0$. The internal flow, as given in Eq.\ref{eq:vminus}, corresponds to rigid body rotation
\begin{equation}
{\bf v}^-({\bf r})=\sum_{m}
c_m^-\frac{r}{a}\mb{e}_r\times {\bf \nabla}_sY_{1m}(\theta, \varphi)={\pmb {\omega}}\times{\bf r}
\end{equation}
with the rotational velocity given by
\begin{equation}
\pmb{\omega}=\frac{1}{3\eta^+}\sqrt{\frac{3}{2\pi}}
\big({\Real}(\gamma_1)\mb{e}_x-{\Imag(\gamma_1)}\mb{e}_y
-\frac{\gamma_0}{\sqrt{2}}\mb{e}_z\big)
\end{equation}

The total force $\mb{F}$ and torque $\mb{N}$ on the droplet
can be
computed from
\begin{equation}
\label{total_force}
\mb{F}= \int_{\Gamma(t)}d^2S\; \mb{t}_{act}, 
 \qquad \mb{N}= \int_{\Gamma(t)}d^2S\;
\mb{r}\times\mb{t}_{act}
\end{equation}
and hence are independent of $\lambda$.
We find
\begin{eqnarray}
\label{explicit_force}
\mb{F}/a^2&=&\sqrt{\frac{4\pi}{3}}(\alpha_0+2\beta_0)\mb{e}_z
+\sqrt{\frac{8\pi}{3}}\Real(\alpha_1+2\beta_1)\mb{e}_x
-\sqrt{\frac{8\pi}{3}}\Imag(\alpha_1+2\beta_1)\mb{e}_y\\
\label{explicit_torque}
\mb{N}/a^3&=&-\sqrt{\frac{4\pi}{3}}2\gamma_0\mb{e}_z
+\sqrt{\frac{8\pi}{3}}2\Real(\gamma_1)\mb{e}_x
-\sqrt{\frac{8\pi}{3}}2\Imag(\gamma_1)\mb{e}_y=8\pi\eta^+{\pmb{\omega}}
\end{eqnarray}

\subsection{Force and torque free case}
\label{squirmer-droplet}

In this section we discuss the special case that no external force and
no external torque are present as required for an autonomous
swimmer. We first work out the consequences for self propulsion by
tractions and then compare our results to the squirmer model in which
an active velocity on the surface of a particle is described. The
squirmer model has been discussed extensively in the literature,
mainly for a solid particle.

Vanishing of the torque (Eq.\ref{explicit_torque}) implies that all
$\gamma_m=0$ and hence the angular velocity vanishes. In other words,
tractions cannot generate rotational motion of an autonomous swimmer.
Vanishing of the total force (Eq.\ref{explicit_force}) implies a
restriction on the traction forces, namely
$\alpha_m=-2\beta_m$. Consequently the Stokeslet vanishes,
i.e. $a_m^+=0$, and the exterior flow field  with $\ell=1$ decays as $1/r^3$.

The flow fields outside and inside of the
force-free droplet are given by
\begin{eqnarray}
\mb{v}^+(\mb{r})&=&-\frac{2a}{3\eta^+(2+3\lambda)}
\sum_{m=-1}^{1} \beta_m\frac{a^3}{r^3}\big (Y_{1m}\mb{e}_r
 -\frac{1}{2}\nabla_s Y_{1m}\big )\nonumber\\
\mb{v}^-(\mb{r})&=&-\frac{a}{3\eta^+(2+3\lambda)}
\sum_{m=-1}^{1} \beta_m\Big((5-3\frac{r^2}{a^2})Y_{1m}\mb{e}_r
 +(5-6\frac{r^2}{a^2})
\frac{1}{2}\nabla_s Y_{1m}\Big )\nonumber
\end{eqnarray}
and the center of mass velocity follows from
Eqs.\ref{vCM_tractions1},\ref{vCM_tractions2} or directly from the constant
part of the velocity field in the interior of the droplet:
\begin{equation}
\mb{v}_{CM}=-\frac{a}{\sqrt{3\pi}(2\eta^++3\eta^-)}\Big(\beta_0\mb{e}_z-
\sqrt{2}\Real\beta_1\mb{e}_x+\sqrt{2}\Imag \beta_1\mb{e}_y
\Big)
\end{equation}

Instead of considering the force-free droplet, one can ask what is the
force needed to keep the active droplet at rest in the LF? We now set
$\mb{v}_{CM}=0$ and compute the stall force , 
\begin{equation}
{\bf F}_{stall}=\int_{\Gamma(t)}(\pmb{\sigma}^{+}\cdot\mb{n})d^2S.
\end{equation}
As one would expect the stall force is proportional to $\mb{v}_{CM}$
\begin{equation} {\bf F}_{stall}=-\mu \mb{v}_{CM} \quad \mbox{with} \quad
  \mu=\frac{(1+\lambda)}{2\pi a(2\eta^++3\eta^-)}
\end{equation}
The mobility $\mu$ is seen to depend on both - the interior and
exterior viscosities. This reflects the fact that traction forces
generate active surface velocities which depend on both viscosities.
In the limit that the fluid in the interior becomes highly viscous or
even solid-like ($\lim \eta^- \to \infty$), the mobility reduces to
the one for a solid particle: $\mu=\frac{1}{6\pi a\eta^+}$. In the
opposite limit $\lim \eta^- \to 0$, we recover the mobility of a gas
bubble: $\mu=\frac{1}{4\pi a\eta^+}$. In between the mobility
increases monotonically.

We now compare our results to the squirmer model which 
prescribes the active velocity field on the surface of the sphere \cite{Blake_1971}:
\begin{equation}
\mb{v}_{act}(\theta, \varphi)=\sum_{m=-1}^{1}\big (\rho_m  \nabla_s Y_{1m}(\theta, \varphi)
\big )
\end{equation}
There is no radial component to the active velocity, because we
require shape and size of the droplet to be preserved.
The  exterior and interior flow field $\mb{v}^{\pm}(r, \theta, \varphi)$ have to 
fulfill the  following boundary condition in the LF:
\begin{equation}
\label{boundary_conditionLF}
	\mb{v}^+(\mathbf{r})=\mb{v}^-(\mathbf{r})=\mb{v}_{CM}+
\mb{v}_{act}(\mb{r})\quad \text{for}\; \mb{r}\in\Gamma(t).
\end{equation} 	
The center of mass velocity is yet unknown and has to be determined from 
the vanishing of the total force (see Eq.\ref{total_force}).
Solving Stokes equation for $\mb{v}^{\pm}$ with the above boundary
condition and the requirement of vanishing total force, we find
for the center of mass velocity
\begin{equation}
\mb{v}_{CM}=\sqrt{\frac{2}{3\pi}} \big(-\frac{\rho_0}{\sqrt{2}}\mb{e}_z
+\Real \rho_1\mb{e}_x-\Imag \rho_1\mb{e}_y\big)
\end{equation}
The two approaches yield the same $\mb{v}_{CM}$, if we identify the
coefficients of the active tractions with those of the active velocity
according to
\begin{equation}
\rho_m= \frac{a\beta_m}{2\eta^++3\eta^-}
\end{equation}

The rotational motion of the squirmer has been discussed by Pak and Lauga\cite{Lauga2014} and more recently by \citet{Pedley2015}. Only the $l=1$ component of the active velocity
contributes to the rotational velocity of the squirmer and is
equivalent to a rigid rotation of the surface with angular velocity
$\pmb{\omega}$, i.e.
$\mb{v}_{act}(\theta, \varphi)=a\pmb{\omega}\times{\bf e}_r$
The active velocity gives rise to a rotation in the LF frame
which exactly cancels the above rotational component of
$\mb{v}_{act}(\theta, \varphi)$ so that the $l=1$ components of the
external and internal flow field are not affected. 
Rotational components with $l>1$ in $\mb{v}_{act}$ do modify the flow
fields inside and outside the droplet, but do not contribute to
$\pmb{\omega}$.

\subsection{Energy Dissipation}
Energy is dissipated by both, the exterior and the interior flow:
\begin{equation}
\partial_tE^{\pm}=\int_{V^{\pm}}d^3x\sigma_{ik}^{\pm}\partial_kv_i^{\pm}=\pm \int_{\partial V}d^2x \;v_i^{\pm}\sigma_{ik}^{\pm}\;n_k
\end{equation}
where we have used that $\partial_k\sigma_{ik}^{\pm}=0$ and the normal
is always defined as pointing outward of the sphere. The second
inequality in the above equation just says that the energy dissipated
in the volumes $V^{\pm}$ is balanced by the energy inflow from the
boundary. The latter is more easily computed and we obtain for the
force free case
 \begin{eqnarray}
\partial_tE^{+}&=&-\frac{4\eta^+a^2}{(2\eta^++3\eta^-)^2}(\beta_0^2+2|\beta_1|^2)\\
   \partial_tE^{-}&=&-\frac{6\eta^-a^2}{(2\eta^++3\eta^-)^2}(\beta_0^2+2|\beta_1|^2)
\end{eqnarray} 
so that the ratio is independent of the driving amplitudes $\{\beta_m\}$:
\begin{equation}
\partial_tE^{+}/\partial_tE^{-}=\frac{2\eta^+}{3\eta^-}
\end{equation}
In the stationary state the total dissipation is balanced by the
energy input due to surface tractions:
$\partial_tE^{+}+\partial_tE^{-}=\partial_tE^{act}=\int_{\partial
  V}d^2x \;t_i^{act}v_i$.

\subsection{Sedimenting Active Droplets}

We only consider the simplest case, namely a passive droplet in an
external potential, which is assumed to be continuous across the
interface. For example, if the fluids are not density matched, then
buoyancy forces act on the droplet. These can be represented by a
potential $U=\Delta\rho g(z-a)$ for $z\leq a$, where $\Delta\rho$
denotes the difference in density between the inside and outside
fluids.

In the general case, the potential is expanded in spherical
harmonics $U=const.+\sum_m U_m(r) Y_{1m}$, keeping only $l=1$. In the
interior of the droplet, we then have to solve Stokes equation
\begin{equation}
	\eta^-\nabla^2 \mathbf{v}^-=\nabla p^- - \mb{f}_{ext} 
\label{stokes_body_force}
\end{equation} 
in the presence of an external force $\mb{f}_{ext}= \nabla U$.  The
force derived from a potential can be absorbed by defining an
effective pressure $p_{eff}=p^--U$. The effective pressure has to
fulfill Laplace`s equation and determines the inhomogeneous solution
of Stokes equation in the same way as does $p^-$ without external
force. The force balance equation for the interface
(eq.\ref{eq:continuity_trac}) involves only the the pressure
$p^-$ and hence the potential $U$ is equivalently represented as a
traction force according to:
\begin{equation}
\mb{t}_{act}(\theta, \varphi)=\sum_{m=-1}^1U_m(a)Y_{1m}(\theta, \varphi)\mb{e}_r
\end{equation}

If there are no active tractions or body forces, we recover the velocity of the sedimenting passive droplet:
$\mb{v}_{CM}=\mu \mb{f}_{ext}$ with
the mobility $\mu=\frac{(1+\lambda)}{2\pi a\eta^+(2+3\lambda)}$.
In the general case of an active droplet in an external field the
motion is a linear superposition of the velocity generated by the
activity and the velocity caused by the external field 
\begin{equation}
\mb{v}_{CM}=-\frac{a}{\sqrt{3\pi}(2\eta^++3\eta^-)}\Big(\beta_0\mb{e}_z-
\sqrt{2}\Real\beta_1 \mb{e}_x+\sqrt{2}\Imag\beta_1\mb{e}_y
\Big)+\mu \mb{f}{ext},
\end{equation}
so that sufficiently strong activity may turn a sedimenting droplet
into a rising one.

\section{Activity generated by body forces}
\label{bodyforces}
So far we have discussed activity which is generated in the interface
of the droplet. We now turn to active mechanisms in the inside of the
droplet. One can either specify active forces or active stresses in
the inside. The latter will be discussed in sec.(\ref{polarmedium}) in the context of a
polar medium which naturally is described in terms of active
stresses. Here we focus on active forces, such as those generated by
motors pulling on filaments and thereby generating flow. Of course,
active forces are related to active stresses by
$\mb{f}^{act}(\mb{r})=-{\bf \nabla}\cdot\pmb{\sigma}^{act}$. However it is not possible
to solve this equation uniquely for $\pmb{\sigma}^{act}$ without using further properties of 
the stress tensor. The situation is analogous to the problem of determinig the electric field $\mb{E}$ from a charge density $\rho$ without using any further properties of $\mb{E}$. Only after specifying that $\nabla\times\mb{E}=0$ the problem becomes well posed. Consequently, we treat the cases of given body force densities and of active stresses separately.

\subsection{A class of body forces}
\label{general_body_forces} 
We start from the $\ell=1$ part of a given force density $\mb{f}^{act}$
\begin{equation}
\mb{f}_{\ell=1}^{act}(r,\theta, \varphi)=\sum_{m=-1}^1\big(\alpha_{m}(r)Y_{1m}(\theta, \varphi)\mb{e}_r
+\beta_{m}(r){\bf \nabla}_sY_{1m}(\theta, \varphi)
+ \gamma_{m}(r) {\bf e}_r\times {\bf \nabla}_sY_{1m}(\theta, \varphi)
\big),
\label{eq:bodyforcedensity}
\end{equation}
which is the analogue of Eq.(\ref{eq:general_traction}). In the
droplet's interior Stokes equation becomes inhomogeneous
\begin{equation}
\label{Stokeswithbodyforce}
\eta^-\nabla^2 \mathbf{v}^-=\nabla p^- - \mb{f}^{act}. 
\end{equation}
Incompressibility requires $\Delta p^-={\bf \nabla}\cdot\mb{f}^{act}$. The
general solution is given as $p^-=p^-_{hom}+p^-_{inh}$ as the sum of
the general solution, $p^-_{hom}$, of the homogeneous equation and a
particular solution, $p^-_{inh}$, of the inhomogeneous equation. The
former has already been constructed in sec.\ref{activity}.  Similarly, the solution
of Eq.\ref{Stokeswithbodyforce} is decomposed into the previously
found flow field for vanishing body force, $\mathbf{v}^-_0=\mathbf{v}^-(\mb{f}^a=0)$
and a particular solution, $\mb{w}$ of the inhomogeneous equation
\begin{equation}
\label{inhStokes}
\eta^-\nabla^2 \mathbf{w}=\nabla p^-_{inh}-\mb{f}^{act} 
\end{equation}

To obtain explicit expressions for $p_{inh}$ and $\mb{w}$ we treat a
class of body forces, $\mb{f}^{act}$, consisting of linear combinations of power laws
$\mb{f}^{act}(r,\theta,\phi)= \sum_\nu (r/a)^{\nu} \mb{f}_\nu(\theta,
\varphi)$,
which may be adapted to specific situations.  Due to the linearity of
the problem, we can restrict the discussion to a single power, so that
$\alpha_m(r)=(r/a)^\nu\alpha_m$, $\beta_m(r)=(r/a)^\nu\beta_m$,
$\gamma_m(r)=(r/a)^\nu\gamma_m$. More general solutions are then
obtained by superposition. 

For $\nu\neq 0$ a special solution of the inhomogeneous equation for the pressure, $\Delta p^-_{inh}={\bf \nabla}\cdot\mb{f}^{act}$, is given by 
\begin{equation}
\label{inh_pressure}
p_{inh}(\mb{r})=(\frac{r}{a})^{\nu+1}\sum_m p_m Y_{1m} \qquad \text{with} \qquad p_m= a \frac{(\nu+2)\alpha_m-2\beta_m}{\nu(\nu+3)},
\end{equation}
We substitute this solution into Eq.\ref{inhStokes} and solve for $\mb{w}$:  
\begin{equation}
\left(\frac{r}{a}\right)^{-\nu-2}\mb{w}=\sum_m w_{m}\left(Y_{1m}\mb{e}_r+\frac{\nu+4}{2}\nabla_s Y_{1m}\right)+u_{m}\mb{e}_r\times\nabla_s Y_{1m},
\label{eq:wsolution}
\end{equation} 
where the coefficients are given by
\begin{eqnarray}\label{inh_velocity}
	w_{m} & = &  \frac{a^2}{\eta^-}\frac{(\nu+1)p_m/a-\alpha_m}{(\nu+2)(\nu+5)}\\\nonumber
		u_{m} & = & -\frac{a^2}{\eta^-}\frac{\gamma_m}{(\nu+1)(\nu+4)}.
	\end{eqnarray}

        After a special solution $\mb{w}$ has been found, the
        coefficients 
        appearing in the homogeneous solution $\mb{v}^+$ and
        $\mb{v}^-_0$ (Eqs.\ref{eq:vplus} and \ref{eq:vminus}) are
        obtained by matching the boundary conditions at $r=a$,
        requiring continuity of the velocity
        Eq. (\ref{eq:continuity_v}), ${\mb v}^+={\mb v}^-_0+{\mb w}$ and
        balance of tractions (\ref{eq:continuity_trac}).  The latter
        requires to calculate the viscous tractions
        $\left(\frac{r}{a}\right)^{-\nu-1}\mb{t}^w=\sum_m \left(
          \epsilon_{m} Y_{1m}\mb{e}_r+\delta_{m}\nabla_s
          Y_{1m}\right)+\zeta_{m}\mb{e}_r\times\nabla_s Y_{1m},$ due to
        the flow field $\mb{w}$:
\begin{eqnarray}
\label{inh_tractions}
	\epsilon_{m}& = & -p_m+2\eta^-(\nu+2)(w_{m}/a)\\\nonumber
	\delta_{m}& =& \eta^- (\nu+2)(\nu+3)(w_{m}/2a)\\\nonumber
	\zeta_{m}& =& \eta^- (\nu+1)(u_{m}/a)
	\end{eqnarray}
       With Eqs. 
        (\ref{eq:vminus}) and (\ref{inh_tractions}) the
        boundary conditions become a linear system of equations, which
        takes on the simple form
\begin{eqnarray}
\label{eq:general_boundcond}
	a^{+}_{m}+b^{+}_{m} -a^{-}_{m}-b^{-}_{m}& = & w_{m}\\\nonumber
	a^{+}_{m}/2 - b^{+}_{m}/2-2a^{-}_m -b^{-}_{m} & = &(\nu+4) w_{m}/2\\\nonumber
	-3\eta^+(a^+_m+2b^+_m)+6\eta^-a^-_m & = & \epsilon_{m}a\\\nonumber
	3\eta^+b^+_m-3\eta^-a^-_m &= & \delta_{m}a
	\end{eqnarray}
	and
\begin{eqnarray}\label{eq:cm}
		c^+_{m}-c^-_{m} & = & u_{m}\\\nonumber
		-3\eta^+c^+_{m}& = &\zeta_{m}a
		\end{eqnarray}

\subsection{Force-free translations}
For the autonomous swimmer, we require that the total force and the total torque of the force density given by Eq.(\ref{eq:bodyforcedensity}) vanishes:
$\mb{F}=\int \mb{f}(\mb{r})\, dV=0$ and 
$\mb{N}=\int \mb{r}\times \mb{f}(\mb{r})\, dV=O$.
If $\alpha_m(r)=-2\beta_m(r)$, the forces within each radial shell $[r, r+dr]$ are balanced.   
Let us specialize to this simple case, with
$\mb{f}=(r/a)^{\nu}\mb{f}_\nu$ and $\alpha_m=-2\beta_m$. $\mb{N}=0$ is
enforced by choosing $\gamma_m=0$.  The last two of the
Eqs.(\ref{eq:general_boundcond}) imply that the amplitudes of the
Stokeslet are given by
$a^+_m=-(1/3\eta^+)(\epsilon_{m}+2\delta_{m})$. These coefficients
vanish, as they have to, for the force free case,
$\alpha_m(r)=-2\beta_m(r)$, if no external forces are present. In this
generic case we get
\begin{eqnarray}
	a_m^+ &= & 0\\
	b_m^+&=  & \frac{\nu/3}{g_\nu(2\eta^++3\eta^-)}a^2\alpha_m\\	
	a_m^-&= & -\frac{(\nu+3)/2+\eta^+/\eta^-}{g_\nu(2\eta^++3\eta^-)}a^2\alpha_m\\
	b_m^-&= & \frac{5\nu+19+6\eta^+/\eta^-}{(\nu+2)(\nu+5)(2\eta^++3\eta^-)}
\frac{a^2\alpha_m}{6}
	\end{eqnarray}
with $g_\nu=\nu(\nu+5)$. The center of mass velocity can be calculated from Eq. (\ref{vCM}) which yields
\begin{equation}
\label{v_CM_bodyforces}
	\mb{v}_{CM}=\sqrt{\frac{3}{2\pi}}(-\Real (b_1^+)\mb{e}_x+\Imag (b_1^+\mb{e}_y)+b_0^+/\sqrt{2}\mb{e}_z)
	\end{equation}

A balance in each radial shell is not possible for the torque. Therefore, within our class of models, a force density with vanishing total force and torque requires at least a sum of two different powers, $\mb{f}=r^\nu\mb{f}_\nu(\theta,\varphi)+r^\mu\mb{f}_\mu(\theta,\varphi)$.

\subsection{Chirality}
As a simple example of a rotating, torque-free droplet, let us consider an active body force density 
$\mb{f}^{act}(\mb{r})=\sum_m\gamma_m(r)\mb{e}_r\times \nabla_s Y_{1m}$ with
\begin{equation}
	\gamma_m(r)=\gamma_{m,1}(r/a)^{\nu_1}+\gamma_{m,2}(r/a)^{\nu_2}.
	\end{equation}
The total torque $\mb{N}$ vanishes, if we require $\gamma_{m,1}/(\nu_1+4)=-\gamma_{m,2}/(\nu_2+4)=\tilde{\gamma}_m$.
The flow field $\mb{v}^-=\mb{v}^-_0+\mb{w}$ inside the droplet takes on the form (see Eq. (\ref{eq:wsolution}))
\begin{equation}
	\mb{v}^-=\sum_{i=1}^2\sum_m \left[c_{m,i}\left(\frac{r}{a}\right)+u_{m,i}\left(\frac{r}{a}\right)^{\nu_i+2}\right](\mb{e}_r\times \nabla_s Y_{1m})
	\end{equation}
with coefficients $c_{m,i}$ and $u_{m,i}$, which are easily obtained from 
Eqs. (\ref{inh_velocity}),(\ref{inh_tractions}) and 
(\ref{eq:cm}.
The resulting flow field is given by   
\begin{equation}
	\mb{v}^-=-\frac{a^2}{\eta^-}\sum_{i=1}^2\sum_{m=-1}^1(-1)^i 
\frac{\tilde{\gamma}_m}{\nu_i+1}\left[\frac{r}{a}-\left(\frac{r}{a}\right)^{\nu_i+2}\right](\mb{e}_r\times\nabla_s Y_{1m}). 
	\end{equation} 
        The flow field $\mb{v}^+$ in the ambient fluid vanishes
        exactly, as it does for the case of a squirmer
        particle \cite{Lauga2014}, but the internal flow 
gives rise to an nonzero angular momentum of the droplet which
        can be computed
        using equ. (\ref{angular-momentum}):
\begin{equation}
	\pmb{\omega}=\frac{c_{\gamma}}{a^3 \eta^-}\sqrt{\frac{3}{2\pi}}[-\mb{e}_x \Real(\tilde{\gamma}_1)+ \mb{e}_y \Imag(\tilde{\gamma}_1)+ \mb{e}_z\tilde{\gamma}_0/\sqrt{2}]
	\end{equation}
with $c_\gamma=\sum_i\frac{(-1)^i}{\nu_i+6}$.

As an illustrating example, consider a chiral, self-propelling droplet with an $\ell=1$ component of body force density given by  
\begin{equation}
	 \mb{f}^{\ell=1}= \sum_m\alpha_mr({Y}_{1m}\mb{e}_r- r\nabla{Y}_{1m}/2)+\gamma_m(5r-6r^2)\mb{e}_r\times \nabla_s{Y}_{1m}
	\end{equation}
	 \begin{figure}
		\includegraphics[width=0.25\textwidth]{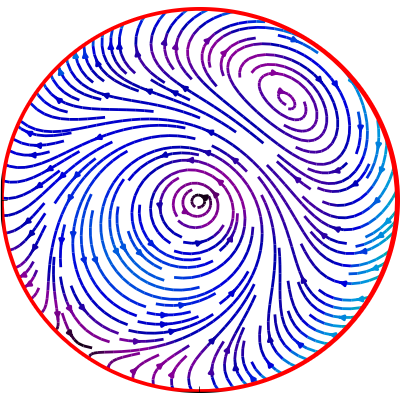}\hfill
		\includegraphics[width=0.25\textwidth]{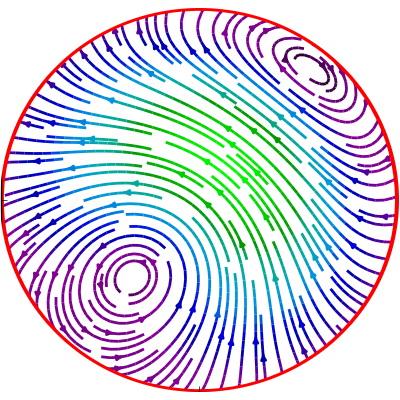}\hfill
		\includegraphics[width=0.25\textwidth]{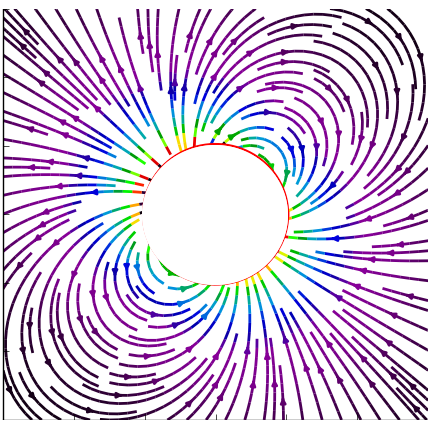}
		\caption{\label{fig:example}The panels show flow lines of the force density and flow fields in the xz plane with parameters $\alpha_0=1, \alpha_1=-\alpha_{-1}^*=1.7+0.5\,i, \gamma_0=1, \gamma_1= -\gamma_{-1}^*=1.2+1.7\,i$. Right panel: force density, middle panel: internal flow field, right panel: external flow field.}
\end{figure}
The total force and torque generated by this density vanishes. In Fig(\ref{fig:example}), the flow lines of the force density $\mb{f}$ and the flow fields $\mb{v}^\pm$ are shown in the x-z plane

\section{Active stresses}
\label{polarmedium}
To make contact with continuum descriptions of active matter
inside a droplet or cell, let us consider an active stress tensor with
cartesian components
\begin{equation}\label{stress_polar}
	\sigma^{(act)}_{ij}=\left(p_i(\mb{r})p_j(\mb{r})-\frac{1}{3}p^2(\mb{r})\delta_{ij}\right).
\end{equation} 
This stress tensor orginates from an active polar medium described by
the polarization field $\mb{p}(\mb{r})$. Its origin is discussed in
detail in Ref. \cite{Kruse2004,Joanny2007} Such active stresses generate a
viscous flow inside the cell or droplet, which in turn gives rise to
self-propulsion. 

We restrict the
discussion to the simple case of a director field, which only contains
$\ell=0$ and $\ell=1, m=0$ modes \footnote{If we start from an expansion of a general $\mb{p}(\mb{r})$
  into vector spherical harmonics, the expansion of the terms
  quadratic in $\mb{p}$, which determine $\mb{f}$, can be performed
  straightforwardly by standard methods \cite{Edmonds}, but it is tedious and the result is not very
  illuminating. It has been worked out in detail in \cite{Moses1974}.} , i.e. it can be written in the form
\begin{equation}
	\mb{p}(r,\theta, \varphi)=\big(p_1(r)\cos\theta+p_0(r)\big)\mb{e}_r-
p_2(r)\sin\theta\mb{e}_\theta,
	\label{eq:director}
\end{equation} 
 To determine the interior flow, we first
compute the body force density
$\mb{f}=-(\nabla\cdot \mb{p})\mb{p}-(\mb{p}\cdot\nabla)\mb{p}
+(1/3)\nabla \mb{p}^2$ with the explicit result given in
Appendix~\ref{appendix_polar}. The polarization of
Eq.~\ref{eq:director} gives rise to force components with
$\ell=0,1,2$. The $\ell=0$-component can be absorbed into the
inhomogeneous pressure (see Eq.(\ref{inh_pressure})) and contributes
to the balance of uniform outside and inside pressure by the uniform
surface tension. We will comment on the $\ell=2$-component below and
focus first on the $\ell=1$-component which determines the overall
motion of the droplet:
\begin{eqnarray}
\label{polar_force_density_l1}
f_r^{(1)}&=&-\frac{4}{3}(p_0p_1)^{\prime}-\frac{4}{r}p_0p_1+\frac{2}{r}p_0p_2\\
f_{\theta}^{(1)}&=& (p_0p_2)^{\prime} + \frac{3}{r}p_0p_2-\frac{2}{3r}p_1p_0\nonumber
\end{eqnarray}
where $p_i^\prime=dp_i/dr$.

Without further assumptions on $\mb{p}(\mb{r})$, we have 
both, active body forces and active tractions, so that the condition of vanishing total force reads
\begin{equation}
\int_VdV\;{\bf f}^a(r)-\int_{\partial V}d^2S\;\pmb{\sigma}^{act}{\bf n}=0
\end{equation}
With the ansatz of Eq.~\ref{eq:director}, force balance can be checked explicitly by computing the two terms separately:
\begin{equation}
  \int_VdV\;{\bf f}^a(r)=-\frac{8\pi}{3}a^2p_0(a)(2p_1(a)/3+(p_2(a)){\bf
    e}_z=  \int_{\partial V}d^2S\;\pmb{\sigma}^{act}{\bf n}.
\end{equation}

\begin{figure}
	\includegraphics[width=0.45\textwidth]{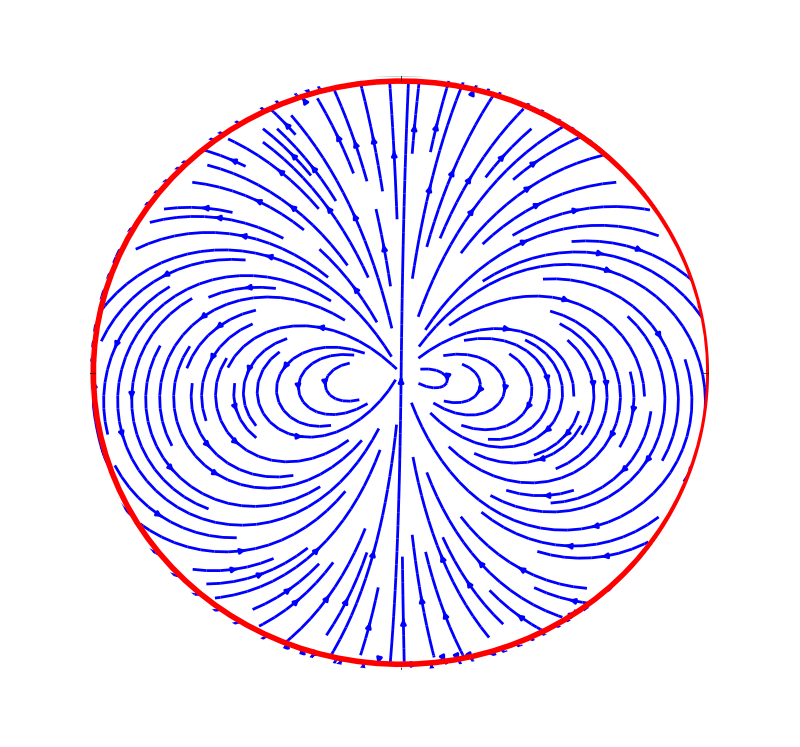}\hspace{2em}
	\includegraphics[width=0.45\textwidth]{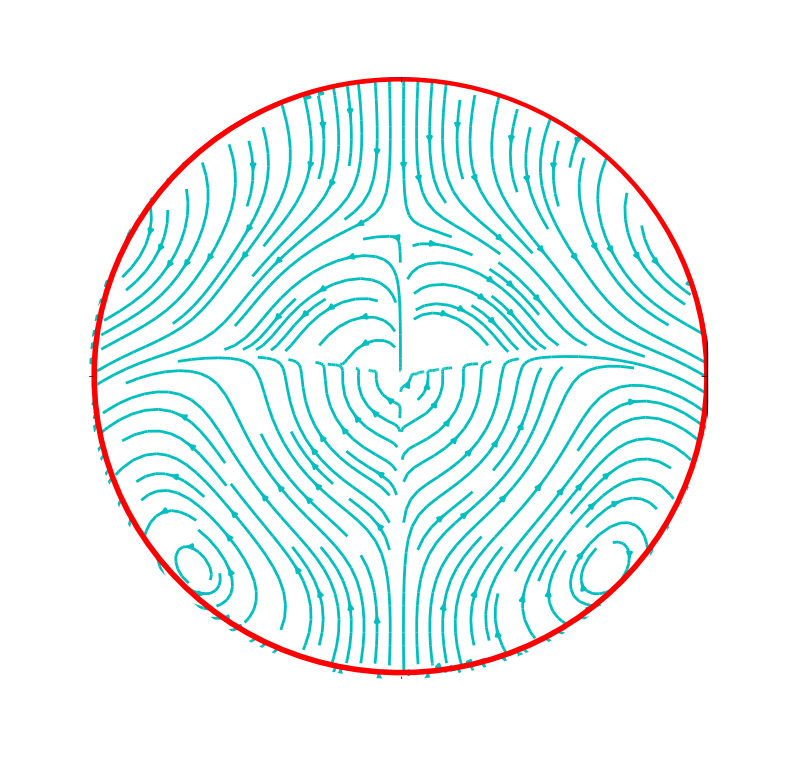}
	\caption{Flow lines in the xz plane. Left: polar order parameter, eq.(\ref{eq:director}), Right: force density, eq.(\ref{eq:forcedensity}) with parameters $p_0=0.2, p_1=p_2=1$}
	\label{fig:polar}
\end{figure}

A simple example for the polarization which vanishes on the surface of
the droplet is: $p_i(r)=p_ir/a (1-r/a)$, implying the vanishing of the
stress tensor on the surface of the droplet and hence the absence of
tractions.  The order parameter field and the corresponding force densities
are shown for this example in Fig.(\ref{fig:polar}).

The external flow field is easily computed from the general formalism
in sec.\ref{general_body_forces}. The details of the calculation are
given in the Appendix~\ref{appendix_polar}. The center of mass
velocity of the droplet is determined by the viscosities of the inside
and outside fluid, $\eta^{\pm}$ and by the coeffcients of the
polarization, $p_0,p_1,p_2$, according to:
\begin{equation} 
{\bf v}_{CM}=\sqrt{\frac{3}{2\pi}}\frac{a p_0}{3(2\lambda+3)}
\big\{\frac{7p_1+p_2}{210\eta^-}-\frac{2p_1+11p_2}{60\eta^+}
\big\}\;{\mb e}_z
\end{equation}

If we think of a nematic liquid crystal, the director field has to
fulfil boundary conditions on the surface of the droplet. We mention
two possibilities: First, the director field is aligned with the
surface, implying $p_0(a)=p_1(a)=0$. Second, the director field is
oriented perpendicular to the surface, implying $p_2(a)=0$. In the
first case, the tractions vanish, 
whereas in the second case, we have both active tractions and active
body forces. The computations for the general case with both, nonzero
body-forces and nonzero tractions is cumbersome but
straightforward. Since we are dealing with linear equations the flow
fields due to tractions as discussed in sec.\ref{activity} and the
flow fields due to body-forces only as discussed in
sec.\ref{bodyforces} can be superimposed.

Before closing this section, two remarks are in order. First, note
that the $\ell=1$ components of the force density are all proportional
to $p_0$, i.e. a director field, which only possesses $\ell=1, m=0$
components, as for example $\mb{p}\propto \mb{e}_z$, will not lead to
self-propulsion. Second and more important, note that our simple
ansatz for the active stress in general gives rise to a force density
with $\ell=2$. ( For the explicit expression, see
Eq.~\ref{polar_force_density} in Appendix~\ref{appendix_polar}). Such a force
density will in general cause deformations of the droplet away from
its spherical shape. Physically these deformations are controlled by
the surface tension, so that it should be possible to describe deformed droplets in a perturbation expansion in $1/\gamma_0$.

\section{Droplet trajectories}
\label{sec:trajectories}
\subsection{Rigid active matter and droplet velocities}
In the previous sections we have calculated the flow fields of a
droplet in the laboratory frame at a single, fixed time. 
Now we we want to construct time dependent flows of the fluids and trajectories of the droplet as a continuum of such single time snapshots.
This is possible if --- in the limit of vanishing Reynolds number --- the Strouhal number $S=a\Omega/U$ stays finite \cite{KimKarrila2005}. Here $U$  is a velocity scale of the internal flow $\mb{v}^-$ , and $S$ compares a time-scale $a/U$ of flow to a time scale $\Omega^{-1}$ of temporal changes of the active forces. 
But the time dependences of active force densities and tractions in the laboratory frame have not yet been specified. These densities are generated by active material, which shares the droplet's interior with the Newtonian fluid, and which possesses its own evolution equations.
The material in the interior of microorganisms has been modelled as  mixtures of simple
 \cite{Dembo1986}  and complex \cite{Kruse2005, Joanny2007} active and passive fluids, or as a biphasic poroelastic medium \cite{Biot1955, deGennes1976, Charras2013} with a distribution of active force densities or
stress tensors, produced by molecular motors and
(dis)-assembly of semi-flexible polymers \cite{Loisel1999}. The resulting evolution equations are quite complex. 
In the present subsection we discuss the case of rigid active matter, immersed in a Newtonian fluid. This model, although highly simplified, is not unreasonable as an approximate continuum description on timescales, which do not exceed the reorganisation time of the cytosceleton. Its main advantage is that it allows for simple closed equations of motion of the droplet.  

In every  continuum description, we label an 
element of the active material by $X$.  At an initial time $t=0$ we assume that the element is at
rest in LF at position $\mb{X}_0$ in a quiet fluid, and in the course of time it follows a trajectory
$\mb{X}(t)=\mb{X}(\mb{X}_0,t)$ with velocity $\mb{v}_m(X,t)=d\mb{X}(t)/dt$.  If, for
example, the material elements do not interact and move convectively in the Newtonian fluid,  the material velocity
equals the fluid velocity $\mb{v}_m(X,t)=\mb{v}^-(\mb{X}(t),t)$.  Here we
restrict the discussion to another simple case, inspired by self-propelling microorganisms. 
We suppose that the active
material is tied to parts of a polymeric (cyto)-sceleton, which can be modelled as a rigid structure. Thus it can only undergo rigid body
motions, composed of rotations $\hat{\mb{R}}(t)$ with respect to some material point and translations
$\mb{b}(t)$, i.e. $\mb{X}(t)=\hat{\mb{R}}(t)\mb{X}(0)+\mb{b}(t)$, and thus the corresponding material velocities are given by $\mb{v}_m(\mb{X},t)=\mb{v}_0+\pmb{\omega}(t)\times [\mb{X}(t)-\mb{b}(t)]$ with $\mb{v}_0=d\mb{b}/dt$. 
 
As an additional constraint we require that the translational motion of the sceleton and the translational motion of the droplet coincide, i.e. $\mb{v}_0=\mb{v}_{CM}$. The constraint can be realized in various ways, for example, by anchoring at least four material points of the sceleton (not all in the same plane) within the interface. This fixes   the center of the sphere rigidly to the sceleton, reminiscent of the anchoring of the cytosceleton at the nucleus. The motion of the sceleton is thus restricted to rotations within the droplet.  Without such a constraint, the sceleton could perform translational motions relative to the center of the sphere --- as, for example, an active particle propelling in the interior of the droplet --- which we do not want to discuss here.

Three frames of reference, which are used in the following, are defined with respect to the sceleton. 
At time $t=0$, when each X is placed at its corresponding $\mb{X}_0$, we choose a cartesian system, with its origin at the center of the sphere, $\mb{r}_{CM}$, and unit vectors $\mb{e}_\alpha, \; \alpha=1,2,3$. This is our laboratory frame LF.  The frame at any other time, with its origin at $\mb{X}(\mb{X}_0=\mb{0}, t)$ and rotated unit vectors
$\hat{\mb{R}}(t)\mb{e}_\alpha$ is called the comoving frame CMF. At $t=0$, CMF and LF coincide, i.e.
$\hat{\mb{R}}(0)=\hat{\mb{1}}$ and $\mb{d}(0)=\mb{0}$. Per definition, the sceleton is at rest in the CMF.  
Its structure is described by a continuous density $\rho_p(X)$, carrying a force density $\mb{f}(X,t)$, which may depend upon time.
Another
laboratory frame, with origin and axes coinciding with those of the CMF at a fixed instant $t$ will be called a momentary  material frame MMF in the following. The calculations of Sec (\ref{activity})- (\ref{polarmedium}) have been performed in this frame.

The active matter in the droplet's interior will in general also modify the internal flow field. 
For a droplet filled with a rigid gel and a
viscous fluid, there is an additional permeation force \cite{Callan2011}, which
enters the force balance. This leads to a modified Stokes equation of the form  
$\nabla\cdot\pmb{\sigma}^-=\nabla p
-\Gamma(\rho_p)(\mb{v}-\mb{v}^m)-\mb{f}^{act}-\mb{f}^{ext}$. Here
$\Gamma(\rho_p)$ is is a permeation
coefficient, which depends upon the detailed structure of the gel.  In the following, we will assume that the gel fraction
is small and $\Gamma(\rho_p)\approx \lambda\rho_p\ll 1$, which allows to neglect the permeation force. Then
$\mb{v}^-$ is entirely due to the active and external force
densities, and can be calculated using Eq. (\ref{eq:stokes}).   

The time-dependence of the force
density $\mb{f}(\mb{r}, t)$ and the sceleton density $\rho_p(\mb{r},t)$ produced by active material at position $\mb{r}$ are now easily obtained. In the CMF, the position and orientation of each material element is time independent and only the explicit time dependence of the force density reamains, i.e. $\rho_p(\mb{r},t)=\rho_p(\mb{X}_0=\mb{r})$ and  $\mb{f}(\mb{r},t)=\mb{f}(\mb{X}_0=\mb{r},t)$. In the LF these quantities are
 co-translated and co-rotated along the trajectory of the active material element, so that 
\begin{equation}
	\label{eq:forceLF}
	\frac{df_i(\mb{r},t)}{dt}=\Big(\mb{v}_{CM}+\pmb{\omega}\times\mb{r}\Big)\cdot\nabla \mb{f}_i(\mb{r},t)+\frac{\partial f_i}{\partial t},
	\end{equation}
and correspondingly $
	d\rho_p(\mb{r}, t)/dt=\Big(\mb{v}_{CM}+\pmb{\omega}\times\mb{r}\Big)\cdot\nabla\rho_p(t).
	$ Thus in the MMF the time dependences of $\rho_p$ and $\mb{f}$ are the same as in the CMF for rigid active matter, which is the main reason for us to use this frame  in our calculations.  
	    
In the following, we will refer to $\pmb{\omega}$ of the sceleton as the angular velocity of the droplet. If the whole droplet is rotating like a rigid body, this definition coincides with that of a self-propelling particle. But  note that there is no unambiguous definition of  angular velocity of a droplet, if the internal flow field $\mb{v}^-$  does not correspond to rigid body rotations.   

To compute the angular velocity of the sceleton from the flow field $\mb{v}^-$, we consider        
the dissipative forces exerted by the flow on the sceleton in the MMF as proportional to the relative velocity between sceleton and fluid, i.e. proportional to $\int_K \rho_p(\mb{v}_m-\mb{v}^-)\, dV$.  A part of these forces is compensated by the rigid constraints, the other part is the only remaining force on the sceleton, which thus has to vanish. A convenient way to calculate the uncompensated generalized force on the rigid sceleton is to start from the dissipation function \cite{Landau}   $D(\pmb{\omega})= (\zeta/2)\int_K \rho_p(\mb{r})(\mb{v}^m(\mb{r})-\mb{v}^-(\mb{r}))^2 dV$. The friction coefficient $\zeta$ is unimportant in the following. The derivative of $D$ with respect to $\pmb{\omega}$ is the generalized force, which has to vanish. The resulting relation
$\int_K[\mb{r}\times(\mb{v}_{CM}+\pmb{\omega}^m\times\mb{r}-\mb{v}^-)]\rho_pdV=0$ can be written in the form 
\begin{equation}
	\label{eq:omegam}
	\sum_{j=1}^3\mb{I}_{ij}(t)\omega_j(t) = \int_K\big[\rho_p(\mb{r},t) \mb{r}\times(\mb{v}^-(\mb{r},t) - \mb{v}_{CM}(t))\big]_i\,dV
	\end{equation}
Here $\mb{I}_{ij}=\int \rho_p(r^2\delta_{ij}-x_ix_j)$. Note that $\pmb{\omega}$ depends on the structure of the sceleton. If it is spherically symmetric $\mb{I}_{ij}=I\delta_{ij}$ with $I=(8\pi/3)\int_0^a\rho_p(r)r^4$ and we get
\begin{equation}
	\label{eq:omegamsph}
	\pmb{\omega}= \frac{1}{I}\int_K\rho_p(r) \mb{r}\times\mb{v}^-\,dV
	\end{equation}
For a homogeneous $\rho_p(r)=\rho_p$ the constant $I$ becomes $I= (8\pi a^5/15)\rho_p$, whereas for a thin shell 
$\rho_p(r)=\rho_p\delta(r-a)$ it is $I=(8\pi a^4/3)\rho_p$. Note that for inhomogeneous $\rho_p(r)$, there is a relative rotational motion between the internal fluid and the sceleton like that observed, for example, in rotations of a cell nucleus\cite{Kumar2014}.

\subsection{Trajectories}

To determine trajectories
of the droplet in LF,  we can use Eq. (\ref{vCM}) and Eqs. (\ref{eq:omegam}, \ref{eq:omegamsph}).
Due to our assumption that the active material is tied to a sparse rigid gel,
the dynamics  of the sceleton in LF is fully characterized  by the transformation of the  MMF, spanned  by coordinate unit vectors  $\mb{e}_i, \; i=1,2,3$ rooted at the center of the sphere, $\mb{r}_{CM}(t)$, into the cartesian LF spanned by $\mb{a}_i, \; i=1,2,3$. The equations of motion for the droplet thus become
\begin{equation}
	\label{eq:trajec}
	\frac{d\mb{r}_{CM}}{dt}=\sum_{i=1}^3 v_{CM, i}(t)\mb{e}_i(t)
	\end{equation}   
\begin{equation}
	\label{eq:rotat}
	\frac{d\mb{e}_{i}}{dt}=\pmb{\omega(t)}\times\mb{e}_i(t).
	\end{equation} 
Here, we insert $\mb{v}_{CM}$ and $\pmb{\omega}$ as calculated from the internal flow. 

In the following we want to present a few simple, illustrating examples, showing trajectories of the center of mass and the rotational motion of a vector $\mb{b}(t)=\sum_i b_i\mb{e}_i(t)-\mb{r}_{CM}(t)$. In the figures depicted below we choose $b_i=1, i=1,2,3$.  
 
\begin{figure}
\includegraphics[width=0.45\linewidth]{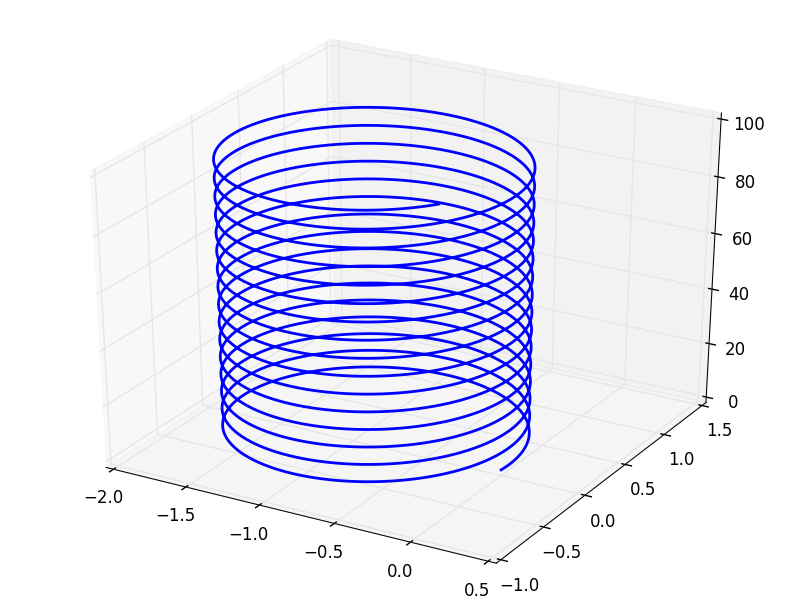}
\includegraphics[width=0.45\linewidth]{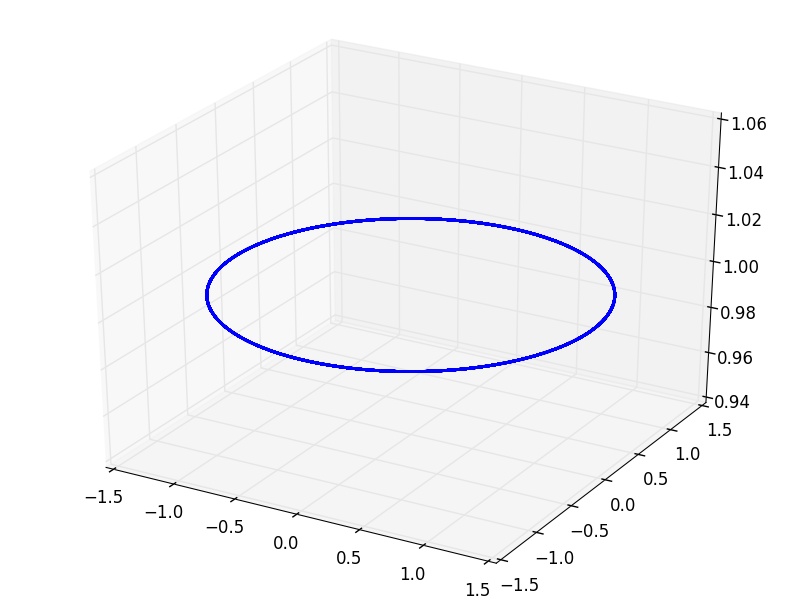}
\caption{\label{fig:timeindependent} For time indendent force densities and/or tractions, the trajectories are straight lines (for $\pmb{\omega}=0$) or  helices (for $\pmb{\omega}\neq 0$). The example shows a trajectory with $\pmb{\omega}=\mb{e}_z$ and $\mb{v}_{CM}=0.3\mb{e}_x+ 0.2\mb{e}_y+\mb{e}_z$. Left panel depicts $\mb{r}_{CM}(t)$, right panel illustrates the rotational motion via $\mb{b}(t)$, defined in the main text.}
\end{figure}

For time-independent force densities, the trajectories from  Eqs. (\ref{eq:trajec}, \ref{eq:rotat}) are helices as shown in Fig (\ref{fig:timeindependent}). The axis of the helix is parallel to $\pmb{\omega}$, which we have chosen to to be the z-axis.       

\begin{figure}
\includegraphics[width=0.45\linewidth]{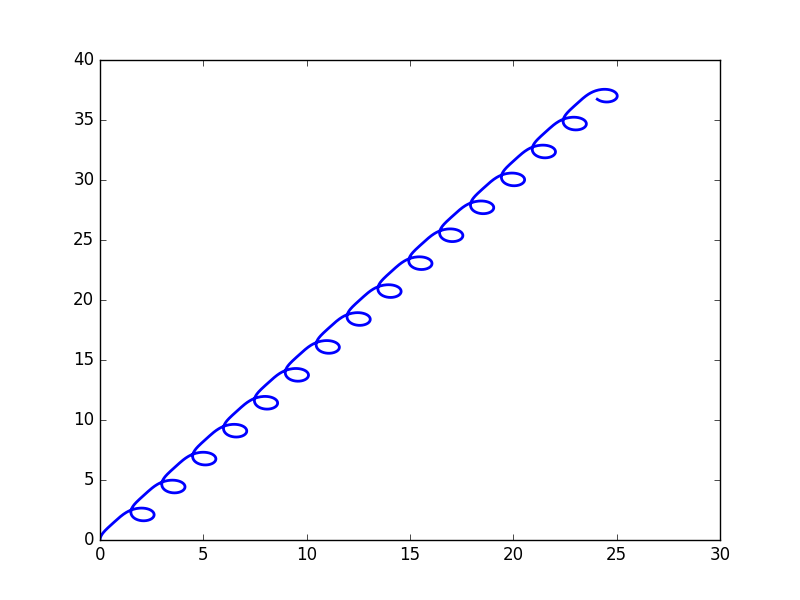}
\includegraphics[width=0.45\linewidth]{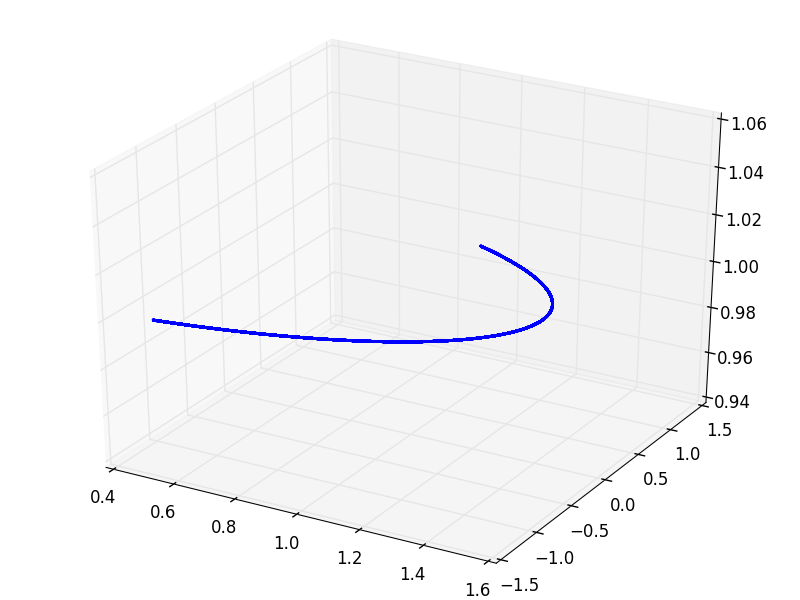}
\caption{\label{fig:periodictranslate} Periodic force densities and/or tractions with vanishing averages over one period may nevertheless lead to unbounded translations. The left panel shows the trajectory of $\mb{r}_{CM}$, the right panel depicts the rotational motion of  $\mb{b}(t)$ (see main text) for $\pmb{\omega}= 0.1\sin(t)\mb{e}_z$ and $\mb{v}_{CM}= 0.1\sin(t)\,\mb{e}_x + 0.1\cos(t)\,\mb{e}_y$}.
\end{figure}

 Fig (\ref{fig:periodictranslate}) shows an interesting case of translational motion. It results from force densities --- and thus from $v_{CM}$ and $\pmb{\omega}$ ---  which vanish when averaged over one period. 
 In the shown example, the angular velocity $\pmb{\omega}=0.1\sin(t)\mb{e}_3$ and the translational velocity $\mb{v}_{CM}=0.1\sin(t)\mb{e}_1+0.1\cos(t)\mb{e}_2$ are simple harmonic functions, and for simplicty are chosen such that the trajectories stay in the x-y plane.  The rotational motion is easily integrated analytically and leads to $\mb{e}_1=\cos\Phi(t)\mb{a}_1-\sin\Phi(t)\mb{a}_2$ and $\mb{e}_2=\sin\Phi(t)\mb{a}_1+\cos\Phi(t)\mb{a}_2$ with $\Phi(t)=1-\cos t$, depicted in Fig.(\ref{fig:periodictranslate} b).

\begin{figure}
\includegraphics[width=0.45\linewidth]{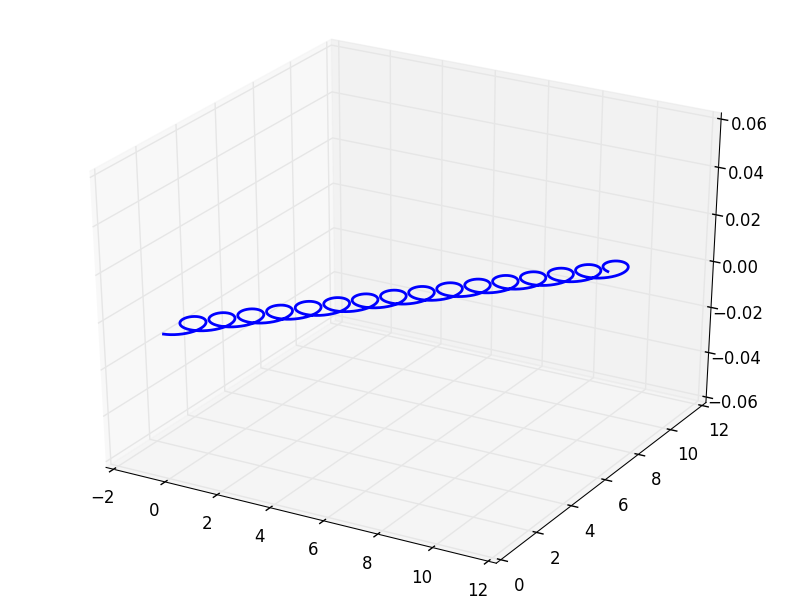}
\includegraphics[width=0.45\linewidth]{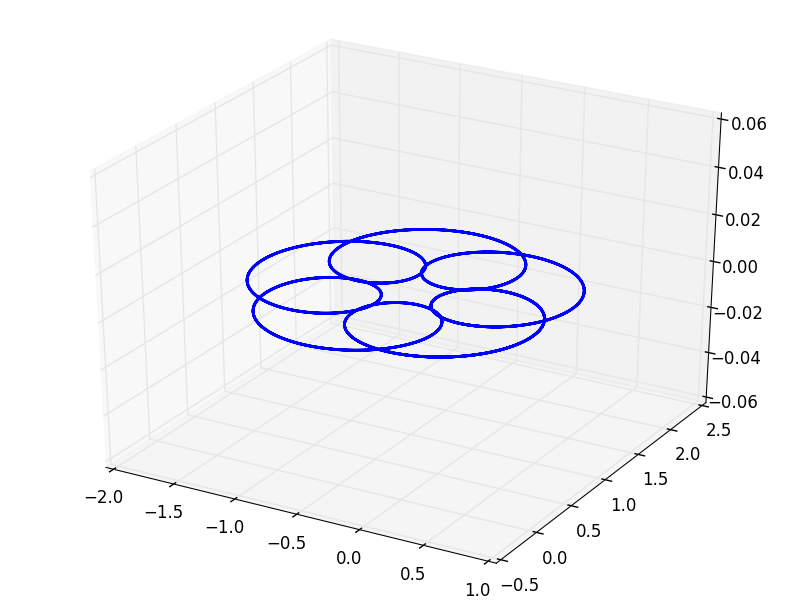}
	\caption{\label{fig:helixwithoutrot} Left panel: Helical motion arising from periodic, non-chiral force densities, which lead to   $\mb{v}_{CM}=(0.1+0.6\cos t)\,\mb{e}_x + (0.1+0.6\sin t)\, \mb{e}_y$ and $\pmb{\omega}=0$. Right panel: A chiral force density, leading to a rotation with  $\pmb{\omega}=0.2\mb{e}_z$ is added to the system of the left panel, resulting in rosette-shaped trajectories. The depicted rosette  is simple and closed, because the the ratio of the frequencies of $\mb{v}_{CM}$ (which is 1) and $\pmb{\omega}$ of this example is a rational number (1/5).}  
\end{figure}

To illustrate the increasing complexity of trajectories, which can be generated, we next add simple cases due to forces with a single frequency $\Omega$ . In Fig(\ref{fig:helixwithoutrot} a), the angular velocity vanishes, nevertheless the  trajectory is a spiral. This type of motion arises from periodic velocities $\mb{v}_{CM}(t)=\sum_i(v_i+u_i\sin(\Omega t))\mb{e}_i$, characterized by a non-vanishing average $\mb{v}=(v_1,v_2,v_3)$ over one period and additional periodic terms of $\mb{v}_{CM}$, which may have different amplitudes $u_i$  $\Omega$ for each cartesian component. 
  
More complex, rosette shaped trajectories --- possibly superimposed with a translation--- appear, if a constant angular velocity is added to periodic translation velocities, as shown in Fig.(\ref{fig:helixwithoutrot} b).

\begin{figure}
\includegraphics[width=0.45\linewidth]{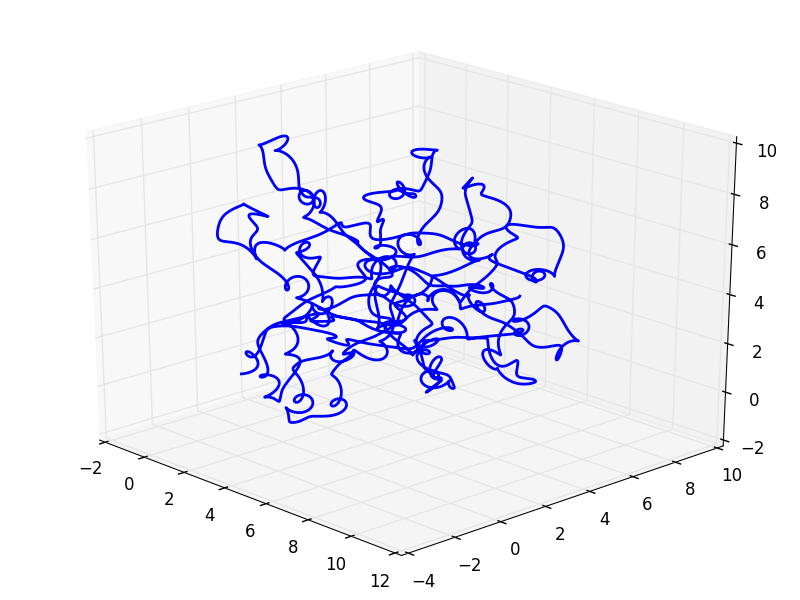}
\includegraphics[width=0.45\linewidth]{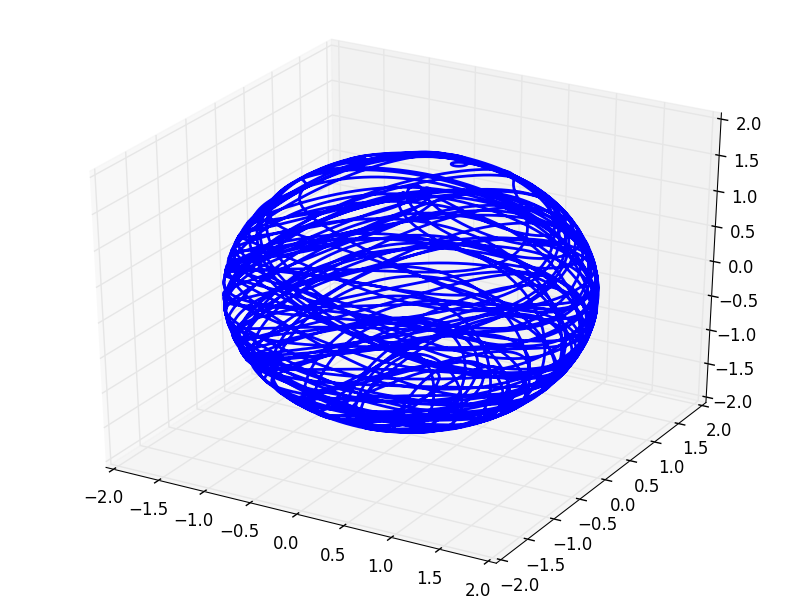}
	\caption{\label{fig:complextrajec} This example shows a complex trajectory of $\mb{r}_{CM}$ and of $\mb{b}$ (see main text), resulting from an angular velocity vector $\pmb{\omega}=0.5\sin t\mb{e}_x+0.3\sin(\sqrt{2} t)\mb{e}_y + \mb{e}_z$ and a translational velocity $\mb{v}_{CM}=0.6\sin(1.9 t)\,\mb{e}_x + 0.8\sin(1.1 t)\,\mb{e}_y + 0.4\sin(0.3 t)\,\mb{e}_z$. It is generated by force densities with 5 different non-vanishing frequencies and a constant component.}
\end{figure}

Finally, if forces contain more than one frequency, the zoology of trajectories gets more and more involved  and will be discussed elsewhere. As an example, Fig. (\ref{fig:complextrajec}) shows the translational and rotational motion of a complex 3-dimensional trajectory with complex rotation.   

\section{Conclusions and Outlook}
\label{conclusions}
We have shown that spherical droplets of immiscible fluids can be
propelled by several activity mechanisms beyond the model of a
squirmer. An important case are active stresses inside the droplet.
An example is an active polar medium, whose polarization ${\bf p}$
generates active stresses of the form
$\sigma_{ij}^{act}\sim p_ip_j-p^2\delta_{ij}/3$. 
Alternatively, the activity is
generated by body forces in the interior of the droplet, an example
are motors pulling filaments and thereby giving rise to a force-dipole
density. Another class of active mechanisms are localised in the
interface of the droplet, giving rise to surface tractions,
$t_j^{act}=n_i\sigma_{ij}^{act}$. These can for example be generated
by active elements in a thin, approximately 2-dimensional
polymer network at the interface.  Active stresses inside the
droplet, which are discontinuous across the interface, also give rise
to tractions localised at the interface, so that the general case
comprises both, tractions and body forces.

Our approach is based on a 2-step procedure.  We first compute the
instantaneous internal and external flow fields as well as the linear
and angular momentum of the droplet. 
The flow fields and the momenta of the droplet are shown to depend on
properties of both fluids, i.e. the viscosities of the interior and
exterior fluid.  Energy is dissipated also in
the interior of the droplet with the ratio of dissipated energies in
the outside and inside being universal, indepedent of the driving
activity.

To compute time dependent flow fields and trajectories of the droplet, we need to specify the
dynamics of the activity mechanism in a second step.  We model these as
being tied to a rigid sceleton, which is constrained to translate with the droplet.  

In the absence of an explicit
time dependence of activity, the droplet trajectories are spirals with in general noncollinear
translational and rotational velocities. Specifically for surface
tractions, as generated e.g. by an inhomogenous surface tension, the
angular velocity vanishes for the autonomous swimmer, i.e. in the
absence of torques.

More generally, we allow for an explicit time-dependence of the
activity mechanism, such as active forces oscillating with frequency
$\Omega$. In order to construct a droplet trajectory as a sequence of
snapshots, which have been computed in the first step of our
procedure, the inertial terms in the Navier Stokes equation have to
remain negligible. This is the guaranteed, if the Reynold`s number is small
and the Strouhal number, $S=a\Omega/U$, remains finite. Here $a/U$ is
a charcateristic timescale of the internal flow. We show that
oscillating active forces can give rise to very complex trajectories,
but postpone a systematic study to future work.

Our approach can be extended in several ways. 
One problem of interest is the motion of an active droplet in nonzero
ambient flow, e.g. a droplet in channel flow or shear flow. This might
involve shape changes of the droplet which have not been considered so
far, because we have mainly restricted ourselves to $l=1$. However a
perturbative analysis of shape changes along the lines of
should be possible \cite{Taylor1934, Vlahovska2009}. The proposed formalism
is general enough to discuss the force free case as well as droplet
motion in the presence of external forces. So far we have analysed the
simplest case only, a sedimenting active droplet, but it is
straightforward to extend the analysis to more complex potentials,
such as an optical trap. Another interesting
extension refers to the interaction of active droplets - possibly
giving rise to collective behaviour.

Finally, the modelling of the droplet can and should be improved in
various ways. One possibility is a two fluid model for the interior,
another line of approach should account explicitly for the gel fraction
and the permeation force. Furthermore it is highly desirable
to study vesicles, enclosed by a membrane, or even cells
enclosed by an elastic cortex. Work along these lines is in progress.

\section{Appendix}
\subsection{General solution of homogeneous Stokes equations in vector spherical harmonics}
To solve boundary value problems of the Stokes equations, 
\begin{equation}
	\eta\nabla^2\mb{v}=\nabla p, \quad \nabla\cdot\mb{v}=0,
	\end{equation}
with a spherical interface $\Gamma(a)$ at $r=a$, we use expansions in terms of scalar and vector spherical harmonics (VSH). Whereas scalar spherical harmonics $Y_{\ell m}$ are fairly standardized, there exist several versions of vector spherical harmonics. Here we use
\begin{eqnarray}
	\label{eq:VSH}
	\mb{Y}_{\ell m 0} & = & Y_{\ell m } \mb{e}_r\\
	\mb{Y}_{\ell m 1} & = & \nabla_s\mb{Y}_{\ell m}\\
	\mb{Y}_{\ell m 2} & = & \mb{e}_r\times\nabla_s\mb{Y}_{\ell m}
	\end{eqnarray}
These functions form a complete orthogonal set on the surface of a sphere with respect to the scalar product
$(\mb{f},\mb{g})=\int_{-1}^{1} d(\cos\theta)\int_0^{2\pi} d\varphi\, \mb{f}^*(\theta, \varphi)\cdot\mb{g}(\theta, \varphi)$. The $s=0$ functions are normalized to $1$, whereas the $s=1,2$ functions have norm $\sqrt{\ell(\ell+1)}$.  

Formulas for the elementary vector analytic operations gradient, divergence  and curl acting on the $\mb{Y}_{\ell m s}$ can be found, for example in \cite{Morse1953}. For the reader's convenience, we include the representation of a gradient and a divergence, which will be used below. For $\nabla p$ one gets
\begin{equation}
	\nabla p(r,\theta, \varphi)=\sum_{\ell m s}(\nabla p)_{\ell ms}\mb{Y}_{\ell m s}=\sum_{\ell, m}\frac{d p_{\ell m}}{dr}\,\mb{Y}_{\ell m 0} +\frac{p_{\ell m}}{r}\,\mb{Y}_{\ell m 1},
	\end{equation}
where $\sum_{\ell m}$ runs over $\ell=0,1,2,\cdots$ and $m$ runs from $-\ell$ to $\ell$. The divergence takes on the form
\begin{equation}
	\nabla\cdot\mb{v}= \sum_{\ell m} (\nabla\cdot\mb{v})_{\ell m} Y_{\ell m}=\sum_{\ell m s} \left(\frac{d v_{\ell m 0}}{dr}+ \frac{2}{r}v_{\ell m 0}-\frac{\ell(\ell +1)}{r}v_{\ell m 1}\right)Y_{\ell m}
	\end{equation}
The incompressibility constraint implies two relations: First, $p$ obeys the Laplace equation  $\nabla^2 p(r,\theta, \varphi)=0$, which is equivalent to 
\begin{equation}
\frac{d^2p_{\ell m}(r)}{dr^2}+\frac{2}{r}\frac{dp_{\ell m}}{dr}-\frac{\ell(\ell+1)}{r}p_{\ell m}=0
 \end{equation} 
 for all possible $\ell$ and $m$, and second the expansion coefficients $v_{\ell m 0}$ and $v_{\ell m 1}$ are related by  
\begin{equation}
	\label{eq:divconstraint}
	\frac{dv_{\ell m 0}}{dr}+\frac{2}{r}v_{\ell m 0}=\frac{\ell(\ell+1)}{r}v_{\ell m 1}
	\end{equation}

The vector Laplace operator $\nabla^2\mb{v}$ is diagonal in $\ell$ and $m$, but for the VSH defined in Eq. (\ref{eq:VSH}) it is not diagonal in the index $s$. It is easily chcked by direct calculation that $\nabla^2\mb{Y}_{\ell m s}=\sum_{s'} h^t_{s's}\mb{Y_{\ell m s'}}/r^2$. The coefficients $h^t_{ss'}=h_{s's}$ define a $3 \times 3$ matrix $\hat{h}$, which is given by
\begin{equation}
	\hat{h}=
	\begin{pmatrix}
		-[2+\ell(\ell+1)] & 2\ell(\ell+1) & 0 \\
		2 & -\ell(\ell+1) & 0 \\
		0 & 0 & -\ell(\ell+1)
		\end{pmatrix},
	\end{equation}  
and which may be used to write the vector Laplace operator in the form
\begin{equation}
	\nabla^2\mb{v}=\sum_{lms} (\nabla^2\mb{v})_{\ell m s}\mb{Y}_{\ell m s} 
	\end{equation}
	with
	\begin{equation}
		 (\nabla^2\mb{v})_{\ell m s}=\sum_{s'} \left\{\left(\frac{d^2}{dr^2}+\frac{2}{r}\frac{d}{dr}\right)\delta_{s s'}+h_{s s'}\right\}v_{\ell m s'}(r)
	 	\end{equation}
The Stokes equations now take on the form of decoupled mode equations,
\begin{equation}
	\label{eq:modeequation}
	\eta(\nabla^2\mb{v})_{\ell m s}=(\nabla p)_{\ell m s}, \quad (\nabla\cdot \mb{v})_{lm}=0.
	\end{equation}
Due to incompressibility the pressure field $p$ has to obey the Laplace equation and thus the $r$ dependence is tight to the $\ell$ modes, leaving the possibilities $p^-\propto r^{\ell}$ for inner and $p^+\propto r^{-\ell-1}$ for outer solutions. Using Eq. (\ref{eq:divconstraint}), $v_{\ell m 1}$ can be eliminated from Eq. (\ref{eq:modeequation}), which is then easily solved. For the resulting inner and out 
modes of the flow fields we find

\begin{eqnarray}
	\label{eq:v+modes}
	v^+_{\ell m0}(r) & = & a^+_{\ell m}(r/a)^{-\ell} + b^+_{\ell m}(r/a)^{-\ell -2}\\ \nonumber
	v^+_{\ell m1}(r) & = & \frac{2-\ell}{\ell(\ell + 1)} a^+_{\ell m} (r/a)^{-\ell} - \frac{1}{\ell+1}b^+_{\ell m} (r/a)^{-\ell -2}\\\nonumber
	v^+_{\ell m2}(r) & = & c^+_{\ell m} (r/a)^{-\ell -1}
	\end{eqnarray}
	and
	 		
\begin{eqnarray}
	\label{eq:v-modes}
	v^-_{\ell m0}(r) & = & a^-_{\ell m}(r/a)^{\ell+1} + b^-_{\ell m}(r/a)^{\ell -1}\\ \nonumber
	v^-_{\ell m1}(r) & = & \frac{\ell+3}{\ell(\ell + 1)} a^-_{\ell m} (r/a)^{\ell+1} + \frac{1}{\ell}b^-_{\ell m} (r/a)^{\ell -1}\\\nonumber
	v^-_{\ell m2}(r) & = & c^-_{\ell m} (r/a)^{\ell},
	\end{eqnarray}
with pressure modes $p^+_{\ell m}(r)=\eta^+a[(4\ell-2)/(\ell + 1)] a^+_{\ell m} (r/a)^{-\ell-1}$ and 
$p^-_{\ell m}(r)=\eta^-a[(4\ell +6)/\ell] a^-_{\ell m} (r/a)^{\ell}$.
The coefficients $a^\pm_{\ell m s}, b^\pm_{\ell m s}$ and $c^\pm_{\ell m s}$ have to be determined from boundary conditions at $\Gamma(r)$ and at infinity. 

To express the boundary conditions on $\Gamma(a)$  we also need the tractions $\mb{t}^{\pm}=\pm\mb{e}_r\cdot\pmb{\sigma}^\pm$, which can be calculated from $\mb{v}^\pm$ and take on the form 

\begin{eqnarray}
	\label{eq:tmodes}
	t^\pm_{\ell m 0}(r) & = & \pm\left[-p^\pm_{\ell m}(r)+ 2\eta^\pm \frac{d v^\pm_{\ell m 0}}{dr}\right]\\\nonumber
	t^\pm_{\ell m 1}(r) & = & \pm\eta^\pm\left(\frac{dv^\pm_{\ell m 1}}{dr}-\frac{v^\pm_{\ell m 1}}{r}+\frac{v^\pm_{\ell m 0}}{r}\right)\\\nonumber
	t^\pm_{\ell m 2}(r) & = & \pm\eta^\pm\left(\frac{dv^\pm_{\ell m 2}}{dr}-\frac{v^\pm_{\ell m 2}}{r}\right)
	\end{eqnarray}  
With the help of Eqs. (\ref{eq:v+modes}), (\ref{eq:v-modes}) and (\ref{eq:tmodes}) the boundary conditions Eqs.(\ref{eq:continuity_v}) and (\ref{eq:continuity_trac}) can now be cast into a system of linear equations. The continuity of flow velocities becomes
\begin{eqnarray}
	\label{eq:abcmodes}
	a^+_{\ell m}+ b^+_{\ell m} -a^-_{\ell m} -b^-_{\ell m}=0\\\nonumber
	(2-\ell) a^+_{\ell m} + \ell b^+_{\ell m} -(\ell +3) a^-_{\ell m}-(\ell +1) b^-_{\ell m}=0\\\nonumber
	c^+_{\ell m}-c^-_{\ell m}=0
	\end{eqnarray}
	and the condition that $\Gamma(a)$ is locally force free takes on the form
\begin{equation}
	\label{eq:t0modes}
	 \eta^+\left(\frac{\ell^2+3\ell-1}{\ell +1}\right)a^+_{\ell m}+\eta^+(\ell+2)b^+_{\ell m} + \eta^-\left(\frac{\ell^2-\ell - 3}{\ell}\right) a^-_{\ell m}+\eta^-(\ell-1)b^-_{\ell m}  = \frac{t^{act}_{\ell m 0}a}{2}
	 \end{equation}
	 \begin{equation}
		 \label{eq:t1modes}
	 -\eta^+\left(\frac{\ell-1}{\ell}\right)a^+_{\ell m} - \left(\frac{\ell+2}{\ell+1}\right)b^+_{\ell m} + \eta^-\left(\frac{\ell +2}{\ell+1}\right)a^-_{\ell m} + \eta^-\left(\frac{\ell-1}{\ell}\right)b^-_{\ell m}=\frac{t^{act}_{\ell m 1}a}{2}
	 \end{equation}
\begin{equation}
	\label{eq:t2modes}
	\eta^+(\ell+2)c^+_{\ell m}+\eta^-(\ell-1)c^-_{\ell m}=\frac{t^{act}_{\ell m 2}a}{2}
	\end{equation}
From Eqs. (\ref{eq:abcmodes} --- \ref{eq:t2modes}) we can now read off the system of linear equations corresponding to the boundary conditions of Eqs. (\ref{eq:continuity_v}) and (\ref{eq:continuity_trac}) for each system of $\ell$-modes.

\subsection{Active polar medium}
\label{appendix_polar}
The force density corresponding to the stress tensor of Eq.~\ref{stress_polar} is of the form
\begin{equation}
\mb{f}=\big(f_r^{(0)}+f_r^{(1)} \cos{\theta}+ f_r^{(2)}(3\cos^2{\theta}-1)
\big)\mb{e}_r+\big(f_{\theta}^{(1)}\sin{\theta}+f_{\theta}^{(2)}\sin{\theta}\cos{\theta}\big)\mb{e}_{\theta}.
\label{eq:forcedensity}
\end{equation}
It thus contains angular momentum components $\ell=0,1,2$, which are explicitly given by:
\begin{eqnarray}
\label{polar_force_density}
-f_r^{(0)}&=&\frac{4}{9}p_1p_1^{\prime}+\frac{4}{3}p_0p_0^{\prime}
-\frac{4}{9}p_2p_2^{\prime}+\frac{2}{3r}p_1^2+\frac{2}{r}p_0^2-\frac{2}{3r}p_2^2\\
-f_r^{(1)}&=&\frac{4}{3}(p_0p_1)^{\prime}+\frac{4}{r}p_0p_1-\frac{2}{r}p_0p_2\\
-f_r^{(2)}&=&\frac{4}{9}p_1p_1^{\prime}+\frac{2}{9}p_2p_2^{\prime}+\frac{2}{3r}p_1^2
-\frac{1}{r}p_1p_2+\frac{1}{3r}p_0^2\\
-f_{\theta}^{(1)}&=&-(p_0p_2)^{\prime}-\frac{3}{r}p_0p_2+\frac{2}{3r}p_1p_0\\
-f_{\theta}^{(2)}&=&p_1p_2^{\prime}-p_2p_1^{\prime}+\frac{3}{r}p_1p_2+\frac{2}{3r}p_1^2
+\frac{7}{3r}p_2^2.
\end{eqnarray}

Let us consider a simple case, for which the polarization vanishes
completely at the interface. We take $p_i(r)=p_ir(1-r)=p_ig(r)$ and compute
the $\ell=1$- components of the force, which are required for the droplet's trajectories:
\begin{eqnarray}
f_r^{(1)}\mb{e}_r&=&-\left(\frac{8}{3}p_0p_1gg'+2p_0\frac{g^2}{r}(2p_1-p_2)\right)\mb{e}_r=\sum_{\nu=1}^3
\alpha(\nu)r^{\nu}\mb{Y}_{10}\\
f_{\theta}^{(1)}\mb{e}_\theta&=&-\left(-2p_0p_2gg'-p_0\frac{g^2}{r}(3p_2-\frac{2}{3}p_1)\right)\mb{e}_\theta=
\sum_{\nu=1}^3 \beta(\nu)r^{\nu}{\bf \nabla}_sY_{10}(\theta).
\end{eqnarray}
For each $\nu=1,2,3$ we solve Eqs.\ref{eq:general_boundcond} and find for the external flow field:
\begin{eqnarray} 
{\bf v}^+({\bf
    r})&=&\sum_{\nu=1}^3b_0^+(\nu)\frac{a^3}{r^3}\big(
  Y_{10}(\theta)\mb{e}_r -\frac{r}{2}{\bf \nabla}Y_{10}(\theta)
           \big)\qquad \text{with}\\
  b_0^+(\nu)&=&-\frac{1}{2(2\lambda+3)}
\big\{\frac{2}{\eta^-}\frac{\alpha(\nu)-(\nu+1)\beta(\nu)}{(\nu+3)(\nu+5)}
-\frac{1}{\eta^+}\frac{\alpha(\nu)+2\beta(\nu)}{(\nu+3)}
                \big\}
\end{eqnarray}
The summation over $\nu$ (done with algebraic computation) then yields
according to Eq.\ref{v_CM_bodyforces} for the center of mass velocity:
\begin{equation} 
{\bf v}_{CM}=-\sqrt{\frac{3}{2\pi}}\frac{p_0}{3(2\lambda+3)}
\big\{\frac{7p_1+p_2}{210\eta^-}-\frac{2p_1+11p_2}{60\eta^+}
\big\}\mb{e}_z.
\end{equation}

\bibliographystyle{jfm}
\bibliography{Microswimmer}

\end{document}